%% file: AlignGPR_arxiv.tex
\documentclass[sigconf,balance=true]{acmart}

\usepackage{multirow}
\usepackage{makecell}
\usepackage{subcaption}
\usepackage{enumitem}

\makeatletter
\IfDocumentMetadataF{%
  \@ifundefined{KV@Gin@artifact}{\define@key{Gin}{artifact}[]{}}{}%
}
\makeatother

\AtBeginDocument{%
  }

\newcommand{\eg}{e.g.,\ }

\makeatletter
\@ifpackageloaded{framed}{}{\RequirePackage{framed}}
\makeatother

\newenvironment{promptbox}{%
  \MakeFramed{\advance\hsize-\width\FrameRestore}%
  \ignorespaces
}{%
  \endMakeFramed
}

\copyrightyear{2027}
\acmYear{2027}
\acmConference[KDD 2027]
  {Proceedings of the 33rd ACM SIGKDD Conference on Knowledge Discovery and Data Mining V.1}
  {August 1--5, 2027}{San Jose, CA, USA}
\acmBooktitle{Proceedings of the 33rd ACM SIGKDD Conference on Knowledge Discovery and Data Mining V.1 (KDD 2027), August 1--5, 2027, San Jose, CA, USA}
\acmISBN{979-8-4007-3134-1/2027/08}

\setcopyright{cc}
\setcctype{by}
\acmDOI{10.1145/3862868.3871360}

\begin{document}

\title[From Category to Venue: AlignGPR]{From Category to Venue: Progressive Category-Collaborative Semantics Aligned Generative POI Recommendation}

\author{Yuxi Huang}
\affiliation{%
  \institution{University of Electronic Science and Technology of China}
  \city{Chengdu}
  \country{China}
}
\email{yuxi.h.cs@gmail.com}

\author{Silin Zhou}
\affiliation{%
  \institution{University of Electronic Science and Technology of China}
  \city{Chengdu}
  \country{China}
}
\email{zhousilinxy@gmail.com}

\author{Wenwen Zhao}
\affiliation{%
  \institution{University of Electronic Science and Technology of China}
  \city{Chengdu}
  \country{China}
}
\email{2023080301028@std.uestc.edu.cn}

\author{Shen Gao}
\affiliation{%
  \institution{University of Electronic Science and Technology of China}
  \city{Chengdu}
  \country{China}
}
\email{shengao@pku.edu.cn}

\author{Shuo Shang}
\authornote{Shuo Shang is the corresponding author.}
\affiliation{%
  \institution{University of Electronic Science and Technology of China}
  \city{Chengdu}
  \country{China}
}
\email{jedi.shang@gmail.com}

\renewcommand{\shortauthors}{Yuxi Huang, Silin Zhou, Wenwen Zhao, Shen Gao, and Shuo Shang}


\begin{abstract}
    Next Point-of-Interest (POI) recommendation is a fundamental task in location-based services. Recently, large language model (LLM)-based generative methods have shown promising progress. However, existing LLM-based methods often overlook POI categories as direct carriers of user mobility intents and neglect collaborative semantics between users and POIs, leading to a disconnect between users' mobility intents and their behavioral patterns. To address this issue, we propose AlignGPR, a progressive collaborative-semantic aligned framework for LLM-based generative POI recommendation, which captures the deep semantic association of \textit{POI category $\rightarrow$ mobility intent $\rightarrow$ behavioral pattern}. AlignGPR introduces \textit{Dual Category-Collaborative Tokenization} to encode category knowledge and collaborative signals into POI and user semantic identifiers (SIDs) via residual quantization. The resulting SID prefixes largely preserve category semantics, while collaborative signals enable fine-grained discrimination for autoregressive generation from user intent to specific POIs. AlignGPR further introduces \textit{Progressive Multi-level LLM Alignment}, which extends the LLM vocabulary with SIDs and progressively aligns the LLM through concept-level index-language grounding, knowledge-level trajectory preference understanding, and task-level generative POI prediction. Experiments on three real-world datasets show that AlignGPR consistently outperforms the strongest baseline, improving MRR by up to 15.99\% with only supervised fine-tuning on a 1.5B LLM.    
    The code is available at 
    \url{https://github.com/yxhuang-Jade/AlignGPR}.
        
\end{abstract}

\ccsdesc[500]{Information systems~Recommender systems}
\keywords{POI Recommendation, Generative Model, LLM}


\makeatletter
\let\AlignGPRsavedfnsymbol\@fnsymbol
\def\@fnsymbol#1{%
  \ifnum#1=1\relax
    \TextOrMath{\textdagger}{\dagger}%
  \else
    \AlignGPRsavedfnsymbol{#1}%
  \fi}
\makeatother
\maketitle
\makeatletter
\let\@fnsymbol\AlignGPRsavedfnsymbol
\makeatother

\input{sections/Intro}

\input{sections/Related_Work_new}

\input{sections/Preliminary}

\input{sections/Methods}

\input{sections/Experimental_Setting}
\input{sections/Conclusion}

\clearpage
\bibliographystyle{ACM-Reference-Format}
\bibliography{ref,model-scale-references}

\appendix

\input{sections/Appendix}

\end{document}

%% file: sections/Intro.tex
\section{Introduction}
\label{Introduction}
\begin{figure}[t]
    \centering
    \includegraphics[width=1\linewidth]{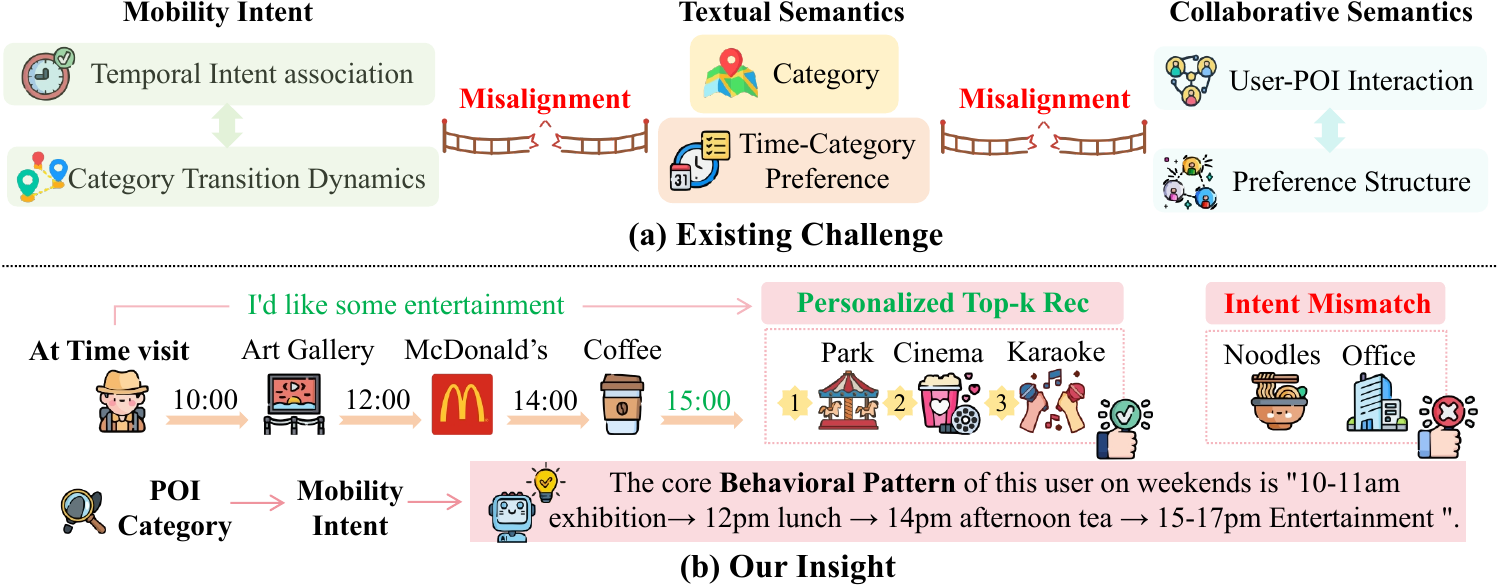}
    \caption{Overview of our motivation. (a) Existing LLM-based POI recommendation methods lack explicit category-intent and textual-collaborative semantic alignment. (b) Illustrates the category-driven behavioral process in user mobility.}
    \label{fig:Intro}
\end{figure}

Next POI recommendation~\cite{Zhang2025POIRec_survey,zhang2022next} aims to predict the next POI a user may visit based on historical check-in records. As a core upstream task in human mobility modeling, it supports a wide range of downstream applications, including trip planning~\cite{Xie2024TravelPlanner}, transport service optimization~\cite{DiDi2026Ride-Hailing}, and route recommendation~\cite{2025Route}. Recently, generative recommendation~\cite{Rajput2023TIGER} has introduced a new paradigm that reformulates recommendation as a sequence generation task, where the model directly generates textual answers, item identifiers, or semantic identifiers rather than scoring candidates. Built upon this paradigm, LLM-based generative recommendation \cite{zheng2024LCRec,Li2024LLM4POI,Wang2025GNPR-SID,kong2025MiniOneRec,Ye2026Align3GR} formulates recommendation as language-conditioned autoregressive generation by serializing item attributes, user behaviors, and task instructions into unified token sequences. This formulation is particularly suitable for next POI recommendation, as the next-visit decision is driven by the joint effect of sequential transition patterns, POI semantics, spatio-temporal contexts, and users' evolving mobility intentions.

Recent LLM-based POI recommendation studies have largely focused on incorporating geographical information into generative POI prediction. For example, LLM4POI~\cite{Li2024LLM4POI} adapts LLMs to check-in histories and POI textual information, while GNPR-SID~\cite{Wang2025GNPR-SID} introduces SIDs for autoregressive POI generation. Subsequent methods further incorporate geographical coordinates~\cite{Liu2025GA-LLM}, POI transition relations~\cite{Zhai2025CoAST}, topology-aware identifiers~\cite{Li2025Refine-POI}, spatial reasoning~\cite{Lv2026Ros}, or bidirectional SID-language alignment~\cite{Zhuang2026Think2Go}. However, these advances remain constrained by the limited fine-grained geographical priors of pre-trained LLMs, making it difficult to reason about complex geographical relationships and leaving performance improvements mostly concentrated on high-frequency toponyms~\cite{Holtermann2025Around,Manvi2024GeoLLM}. More fundamentally, strengthening geographical representations alone does not resolve how recommendation-oriented semantics should be grounded and aligned within an LLM.

However, this recommendation-semantic bottleneck involves two coupled misalignments, as illustrated in Figure~\ref{fig:Intro}(a). 
First, existing methods fail to model POI categories as intent signals, and the inadequate handling of users' temporal category preferences further results in a lack of mobility-intent knowledge at the level of textual semantics.
Consequently, LLMs struggle to establish temporal category-intent associations, capture category transition dynamics, and constrain generation toward intent-consistent POIs. Second, textual semantics weakly align with collaborative semantics, such as user–POI interactions and personalized preferences. Existing approaches generally rely on next-POI prediction to implicitly learn textual–behavior correspondences, yet fail to explicitly model how mobility intents are grounded in trajectory-level behaviors. Thus, LLMs cannot adequately understand how category-level meanings are rooted in user behaviors, nor determine which category-consistent POIs truly match the target user's preferences.

Bridging these two misalignments is essential because \textbf{POI recommendation is inherently category-collaborative.} 
We view category semantics and collaborative signals as tightly connected aspects of intent-driven mobility, rather than independent features:
POI categories reveal users' visit purposes, while user-POI interactions further ground these category-level purposes in implicit and personalized behavioral preferences. 
This perspective differs from a focus on incorporating user-item collaborative signals into LLMs for recommendation~\cite{Zhang2025CoLLM}.
Traditional POI recommendation methods have demonstrated the effectiveness of modeling category-aware behavioral semantics~\cite{Yang2017PACE,He2017category,Rao2025Disentangled}. For example, LBPR~\cite{He2017category} predicts the user's preferred next category before ranking POI candidates, while DPRL~\cite{Rao2025Disentangled} models time and category contexts in a disentangled and personalized manner. These findings are consistent with studies on categorization and decision making~\cite{Wang2016categorization,Zheng2023categorization} and activity-based travel analysis~\cite{hoogendoorn2004pedestrian,rasouli2014activity}, which suggest that users often determine the activity category of a visit before selecting a concrete POI. 
As shown in Figure~\ref{fig:Intro}(b), this category-first decision process underlies the trajectory 
``exhibition $\rightarrow$ lunch $\rightarrow$ afternoon tea $\rightarrow$ entertainment'', which reflects a behavioral progression from \textbf{POI category} to \textbf{mobility intent} and then to \textbf{behavioral pattern}, rather than merely a chronological sequence of visited locations. Without sufficient alignment between category semantics and temporal mobility-intent knowledge, an LLM may fail to infer the user's emerging entertainment intent; even when entertainment is recognized as a plausible next category, weak alignment with collaborative semantics prevents the model from identifying which entertainment POIs best match the target user's personalized preferences. 


To address these issues, we propose Progressive Collaborative Semantics \textbf{Align}ed \textbf{G}enerative \textbf{P}OI \textbf{R}ecommendation, referred to as \textbf{AlignGPR}, which explicitly treats category semantics as dominant mobility-intent signals and couples them with user-POI collaborative semantics throughout tokenization and LLM alignment. AlignGPR consists of two parts: first, \textit{dual category-collaborative tokenization} constructs isomorphic POI and user SIDs by jointly encoding category semantics and collaborative signals through separate residual quantizers. Following a coarse-to-fine principle, category semantics serve as coarse-grained anchors that organize the SID space around mobility intents, while collaborative signals distinguish POIs and users within similar category regions and guide collision resolution for unique identification. Second, \textit{progressive multi-level LLM alignment} incorporates the constructed SIDs into the LLM vocabulary and organizes the associated supervision into a concept-knowledge-task curriculum. At the concept level, bidirectional index-language tasks ground individual POI and user SIDs in their category and preference meanings; at the knowledge level, trajectory modeling and category-intent description tasks teach the LLM to understand temporal-category regularities, mobility-intent evolution, and collaborative preference patterns; at the task level, the LLM is aligned for next-POI prediction. By replacing conventional mixed multi-task fine-tuning with this stage-wise progression, AlignGPR respects the inherent dependency among alignment tasks, alleviates training imbalance, and enables LLMs to progressively acquire recommendation-oriented textual and collaborative semantics by supervised fine-tuning.

We evaluate AlignGPR on three real-world POI recommendation datasets. The results show that AlignGPR consistently outperforms the strongest baseline and exhibits strong generalization in cold-start and cross-domain scenarios. In particular, AlignGPR improves MRR by up to 15.9\% over the strongest baseline using only supervised fine-tuning with a 1.5B LLM.

Our key contributions are summarized as follows.
\begin{itemize}[leftmargin=*]
    \item We introduce AlignGPR, an LLM-based generative POI recommendation framework that explicitly models POI categories as independent information and captures users’ deep semantic correlations across category, mobility intent, and behavioral pattern.
    \item We propose a category-aware bidirectional quantization scheme and elaborately construct fine-tuning tasks rooted in categorical semantics and collaborative preferences, achieving bidirectional coupling between category and collaborative semantics across both tokenization and fine-tuning stages.
    \item We present a progressive multi-level alignment training paradigm. Through experiments, we reveal the importance of semantic initialization for SIDs and the effectiveness of decoupling the semantic alignment task from the main recommendation task.
\end{itemize}

\begin{figure*}[t]
    \centering
    \includegraphics[width=\textwidth]{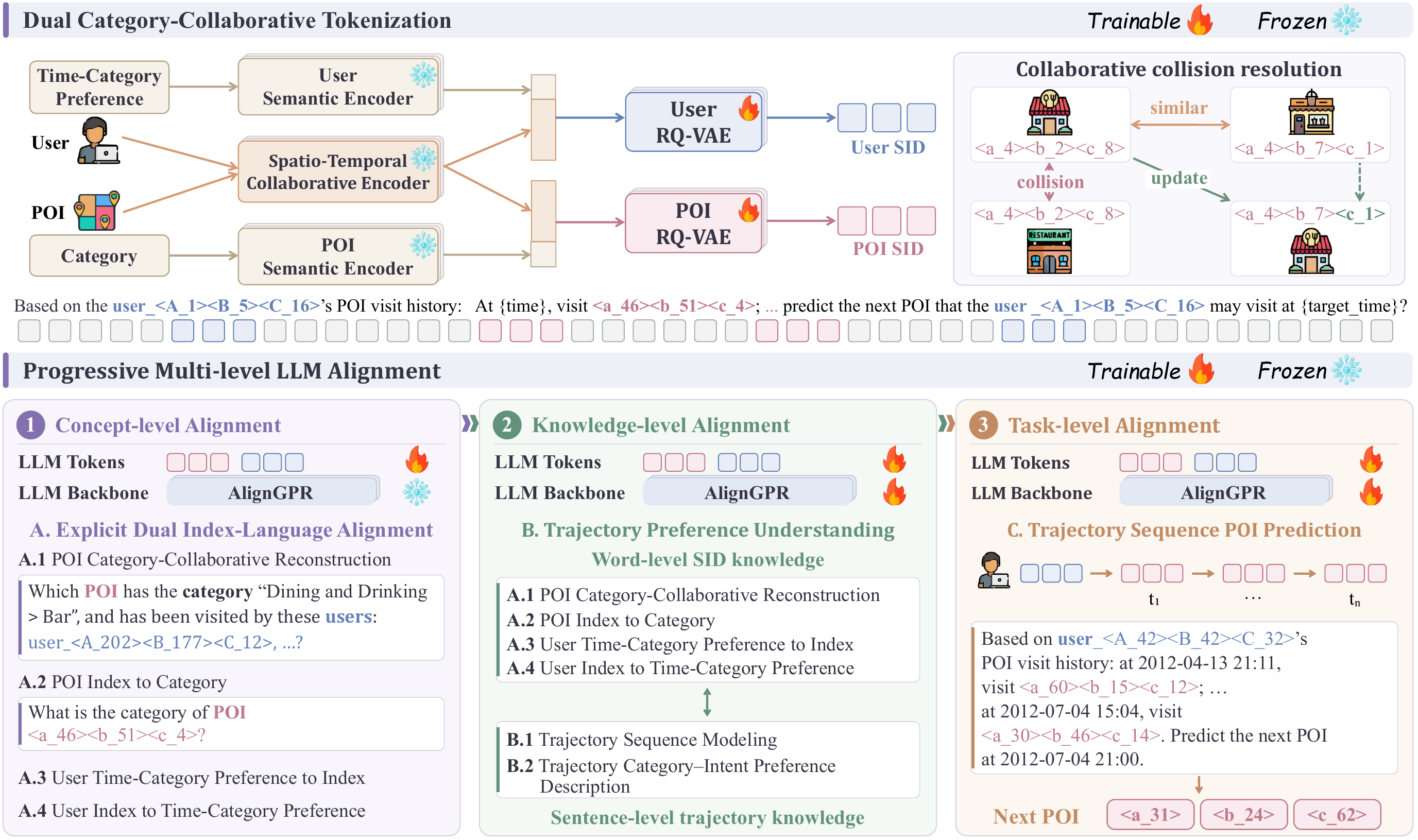}
    \caption{The overall framework of AlignGPR.}
    \label{fig:framework}
\end{figure*}

%% file: sections/Related_Work_new.tex
\section{Related Work}
\label{Related Work}

\noindent\textbf{Model-based POI Recommendation.}
Early POI recommendation studies mainly relied on personalized Markov chain variants~\cite{cheng2013Markov,Feng2015Markov} to model sequential transition patterns. With the development of deep learning, RNN-based~\cite{Liu2016STRNN,Yao2017Serm,Feng2018Deepmove}, Transformer-based~\cite{Luo2021Stan,Wang2022Interval,Xie2023STAR-HiT}, and GNN-based methods~\cite{Scarselli2008Graph,Rao2022Graph-flashback,Yan2023STHGCN} have been proposed to capture sequential, spatio-temporal, and graph-structured user-POI interactions. Most modern model-based methods follow the dense retrieval paradigm~\cite{Yang2025Sparse}, which learns low-dimensional embeddings for users and POIs and ranks candidates by similarity scores. Within this paradigm, POI category information has been recognized as fundamental for understanding user intent~\cite{Tian2023Category,Rao2025Disentangled}, alleviating data sparsity~\cite{Wang2023CCDSA,Zhu2024MGCL}, and capturing transition regularities~\cite{He2017LBPR,Lin2018Category,Li2025GDPW}. For example, CCDSA~\cite{Wang2023CCDSA} jointly models fine-grained POI preferences and coarse-grained category preferences, while GDPW~\cite{Li2025GDPW} constructs category and category-temporal graphs to model time-varying transitions across POI categories. However, this discriminative paradigm still relies on randomly assigned IDs and similarity-based ranking, making it difficult to fully exploit the semantic correlations embodied by users, POIs, categories, and their interactions.

\noindent\textbf{Semantic-based POI Recommendation.}
Grounded in the generative retrieval paradigm~\cite{Li2025GRsurvey}, LLM-based generative recommendation~\cite{zheng2024LCRec,cikm25-gen-rec-tutorial} reframes POI recommendation as an end-to-end sequence generation problem, where the key challenge is aligning behavioral semantics with LLM token semantics through fine-tuning. 
LLM4POI~\cite{Li2024LLM4POI} reformulates next POI recommendation as a question-answering task that uses textual descriptions of user check-in trajectories to predict the next POI ID, while GNPR-SID~\cite{Wang2025GNPR-SID} generates semantically consistent SIDs with similarity priors for autoregressive POI recommendation. Gwhere~\cite{Zhai2025CoAST} pre-trains on SID-location-description alignment and behavior sequence corpora, and ROS~\cite{Lv2026Ros} introduces explicit index-language alignment tasks. Refine-POI~\cite{Li2025Refine-POI} and Think2Go~\cite{Zhuang2026Think2Go} further improve top-$K$ recommendation through explicit reasoning over spatio-temporal regularities in user mobility. However, existing LLM-based approaches still underexplore collaborative semantics and the semantic value of POI categories, limiting their ability to capture user intent from category transitions and mine behavioral patterns for fine-grained personalization.

%% file: sections/Preliminary.tex

\section{Preliminary}
\label{sec:preliminary}

\noindent\textbf{Definition 1 (Check-in).}
Let $U = \{u_i\}_{i=1}^{M}$ and $V = \{v_i\}_{i=1}^{N}$
represent the set of $M$ users and $N$ POIs, respectively. A user check-in is denoted as a tuple $p = (v, c, t)$, indicating that user $u$ visited POI $v$ at time $t$, and $c$ denotes the category of the POI.

\noindent\textbf{Definition 2 (Trajectory).}
A trajectory is a time-ordered sequence of check-in records made by user $u \in U$, denoted as $\tau = \langle p_i \rangle_{i=1}^{n}$, where $p_i$ represents the $i$-th check-in event of user $u$ and $n$ is the length of this trajectory. 

\noindent\textbf{Problem Statement (Generative POI Recommendation).}
Given a user $u \in U$, the historical trajectory $\tau$,
and the target time $t_{n+1}$ of the next visit, the task aims to predict the next POI
$v_{n+1} \in V$ that user $u$ will visit at time $t_{n+1}$. In the generative
recommendation paradigm, each POI $v$ is represented by a sequence of identifiers
$Y_v = [y_1, y_2, \dots, y_L]$, where $y_\ell \in \mathcal{Y}$ and $\mathcal{Y}$
denotes the generation vocabulary. Given an input sequence $X$ consisting of task
instructions, $u$, $\tau$, and $t_{n+1}$, the target POI identifier sequence
$Y_{v_{n+1}}$ is generated directly. The joint generation probability is factorized as:
\begin{equation}
\Pr(Y_{v_{n+1}} \mid X) = \prod_{\ell=1}^{L} \Pr\left( y_\ell \mid X, y_1, y_2, \dots, y_{\ell-1} \right).
\end{equation}

%% file: sections/Methods.tex
\section{Methods}
\label{Methods}

Figure~\ref{fig:framework} presents the overall framework of AlignGPR. First, \textit{Dual Category-Collaborative Tokenization} constructs POI and user SIDs by integrating category semantics and collaborative signals through residual quantization, with collision handling guided by collaborative semantics. Then, \textit{Progressive Multi-level LLM Alignment} extends the LLM vocabulary with the constructed SIDs and organizes them into multi-grained task prompts. Finally, AlignGPR progressively aligns the LLM from concept-level alignment to knowledge-level alignment and task-level alignment toward the final next-POI recommendation objective.

\subsection{Dual Category-Collaborative Tokenization}
\label{sec:sid_tokenization}
In generative item recommendation, standard tokenization directly applies residual quantization to transform item representations into SIDs. 
However, POIs differ from items in that their textual information often mixes heterogeneous semantics: long geographic descriptions include spatial information and mainly encode fine-grained locations and surrounding areas, while short category texts directly reflect mobility intents. Jointly encoding these signals into a representation may bias residual quantization toward geographic semantics. Moreover, since LLMs do not possess fine-grained geographic knowledge, such geography-dominated SIDs receive limited benefit from the semantic knowledge encoded in LLMs, thereby weakening the connection between SIDs and users' travel intentions. We therefore construct category-centered SIDs under a coarse-to-fine principle. Category semantics serve as coarse-grained anchors for grouping POIs and users with similar mobility intents, while collaborative signals provide fine-grained discrimination within similar category regions by encoding user-POI interactions, geographic proximity, and neighborhood structures.

\noindent\textbf{Dual Category-Collaborative Representations.} To establish collaborative associations between users and POI category semantics in the textual space, we first leverage an LLM (\eg Deepseek-v3.2~\cite{Liu2025Deepseek}) to generate a long-term temporal-category preference profile $\mathcal{P}_u^{\mathrm{cat}}$ from each user's historical trajectory.
The profile jointly characterizes the user's preferences for different POI categories, their associations with temporal contexts (\eg time-of-day and weekday/weekend patterns), and how these associations evolve over time. 
Each profile is constructed exclusively from the check-ins observed before the corresponding prediction target, excluding the target check-in and all subsequent interactions. By verbalizing the temporal-category regularities derived from user--POI interactions, these profiles connect user preferences with POI category semantics in natural language and provide user-specific textual grounding for the intrinsic relation of ``POI category $\to$ travel intention $\to$ behavioral pattern''.

We first encode each profile using a frozen semantic encoder (e.g., Qwen3-Embedding-4B~\cite{Zhang2025Qwen3Emb}) to obtain the text embedding sequence, then apply mean-pooling over this text embedding sequence to get the temporal-category profile embedding as follows:
\begin{equation}
\mathbf{e}_u^{\mathrm{cat}}
=
\mathrm{MeanPooling}
\left(\mathrm{SemanticEncoder}
\left(
\mathcal{P}_u^{\mathrm{cat}}
\right)\right) \in
\mathbb{R}^{d}.
\end{equation}

Similarly, for POI $v$, we obtain the textual description of its category, denoted as $\mathcal{P}_v^{\mathrm{cat}}$, and apply the same semantic encoding and mean-pooling process to derive its category embedding:
\begin{equation}
\mathbf{e}_v^{\mathrm{cat}}
=
\mathrm{MeanPooling}
\left(
\mathrm{SemanticEncoder}
\left(
\mathcal{P}_v^{\mathrm{cat}}
\right)
\right)
\in
\mathbb{R}^{d}.
\end{equation}

In parallel, we also employ a frozen spatio-temporal collaborative encoder to capture collaborative signals from user-POI interactions and mobility patterns. Here, we adopt STHGCN~\cite{Yan2023STHGCN}, whose graph-based architecture explicitly models POI geographic proximity, temporal visit patterns, and high-order user-POI collaborative relationships. The collaborative embeddings of POI $v$ and user $u$ are obtained as follows:
\begin{equation}
\begin{aligned}
\mathbf{e}_v^{\mathrm{coll}}
&=
\mathrm{CollaborativeEncoder}(v)
\in
\mathbb{R}^{d'}, \\
\mathbf{e}_u^{\mathrm{coll}}
&=
\mathrm{CollaborativeEncoder}(u)
\in
\mathbb{R}^{d'}.
\end{aligned}
\end{equation}

Finally, we concatenate the category and collaborative embeddings to obtain the final POI and user representations:
\begin{equation}
\mathbf{e}_v
=
\left[
\mathbf{e}_v^{\mathrm{cat}},
\mathbf{e}_v^{\mathrm{coll}}
\right]
\in
\mathbb{R}^{d+d'},
\qquad
\mathbf{e}_u
=
\left[
\mathbf{e}_u^{\mathrm{cat}},
\mathbf{e}_u^{\mathrm{coll}}
\right]
\in
\mathbb{R}^{d+d'}.
\end{equation}

\noindent\textbf{Residual Quantization.}
After obtaining the category-collaborative representations of POIs and users, we map these representations into discrete indices using two independent Residual Vector Quantized Variational Autoencoders (RQ-VAEs). 
This separation accommodates the distinct representation distributions of POIs and users while applying the same hierarchical quantization procedure to both entity types. 
For an input embedding $\mathbf{e}$, which can be either the POI embedding $\mathbf{e}_v$ or the user embedding $\mathbf{e}_u$, the corresponding RQ-VAE first compresses it into a latent embedding $\mathbf{z}$. We set the number of residual quantization layers to $N_q$ (i.e., the depth of the hierarchical SID). At each layer $i \in \{1, \dots, N_q\}$, we define a learnable codebook $\mathcal{B}^{(i)} = \{ \mathbf{b}_{j}^{(i)} \}_{j=1}^{K}$, where each codebook vector serves as a learnable cluster center. The detailed residual quantization process is formulated as:
\begin{equation}
k_i
=
\arg\min_{j \in \{1,\ldots,K\}}
\| \mathbf{r}_i - \mathbf{b}_{j}^{(i)} \|_2^2,
\end{equation}
\begin{equation}
\mathbf{r}_{i+1}
=
\mathbf{r}_i - \mathbf{b}_{k_i}^{(i)},
\end{equation}
where $k_i$ denotes the code index selected at the $i$-th layer, $\mathbf{b}_{k_i}^{(i)}$ denotes its corresponding codebook vector, $\mathbf{r}_i$ denotes the residual vector at the $i$-th layer, and the initial residual vector is set as $\mathbf{r}_1 = \mathbf{z}$. Given $N_q$ stacked codebook layers, the quantized latent representation is reconstructed by $\hat{\mathbf{z}} = \sum_{i=1}^{N_q} \mathbf{b}_{k_i}^{(i)}$.

The quantized vector $\hat{\mathbf{z}}$ is fed into the decoder to recover the original input embedding $\hat{\mathbf{e}}$. The complete objective functions are defined as follows:
\begin{equation}
L_{recon} = \| \mathbf{e} - \hat{\mathbf{e}} \|_2^2,
\end{equation}
\begin{equation}
L_{rqvae}
=
\sum_{i=1}^{N_q}
\left(
\| sg[\mathbf{r}_i] - \mathbf{b}_{k_i}^{(i)} \|_2^2
+
\beta \| \mathbf{r}_i - sg[\mathbf{b}_{k_i}^{(i)}] \|_2^2
\right),
\end{equation}
\begin{equation}
\mathcal{L} = L_{recon} + L_{rqvae},
\end{equation}
where $\hat{\mathbf{e}}$ is the decoded output embedding, $sg[\cdot]$ refers to the stop-gradient operation, and $\beta$ is a hyper-parameter for loss weighting. The total loss consists of two components: the reconstruction loss $L_{recon}$ ensures representation restoration, while the residual quantization loss $L_{rqvae}$ narrows the distance between residual vectors and learnable codebook embeddings. Therefore, the final hierarchical discrete index for an entity $x$ is formalized as:
\begin{equation}
\mathbf{k}_x
=
\left[
k_x^{(1)},
k_x^{(2)},
\ldots,
k_x^{(N_q)}
\right],
\quad
k_x^{(i)} \in \{1, \ldots, K\}.
\end{equation}

For POI and user representations, we adopt two standalone RQVAEs to perform the identical residual quantization pipeline. The generated SIDs (\eg \texttt{<a\_1><b\_2><c\_3>}) serve not only as discrete identifiers, but also as a new language distinct from natural text, enabling LLM-based alignment to build collaborative associations between textual semantics and collaborative semantics.

\noindent\textbf{Handling Collisions via Collaborative Semantics.}
SID is a tree structure derived from residual quantization, which may cause multiple POIs to be clustered into the same leaf node. However, as an identifier, it must satisfy uniqueness---each SID should uniquely correspond to one POI. To address this issue, some methods~\cite{Ju2026Snap,zheng2024LCRec,Lv2026Ros,Wang2025GNPR-SID} append a quantization-independent identifier after the index, resolving the conflict at the cost of disrupting the semantic structure of the SID, which is not an ideal choice.

To resolve SID collisions, we modify only the final SID level while keeping all preceding levels unchanged. It is worth noting that all code indices are derived from the same category-collaborative representation, rather than being separately assigned to category or collaborative semantics. Because the semantic component is constructed exclusively from POI category descriptions and user temporal-category preference profiles, it provides the explicit semantic anchor of the joint representation. In contrast, the collaborative component obtained from user--POI interactions and mobility patterns supplies behavioral proximity signals that category text alone cannot capture. Under this category-centered representation, category semantics provide the basis for coarse SID organization, while collaborative signals refine distinctions among entities within similar category regions. Accordingly, our collision handling preserves all preceding category-collaborative code indices and modifies only the final residual code index for unique identification. For each conflicting entity $x$ (i.e., a POI or a user), we rank entities of the same type by their collaborative similarity to $x$ and replace its final-level code index with that of the highest-ranked candidate whose code index is unoccupied under the prefix defined by $x$'s preceding code indices.

For a conflicting entity $x \in \mathcal{X}$, where
$\mathcal{X}\in\{V,U\}$ denotes either the POI set or the user set,
let $\mathcal{G}_x\subseteq\mathcal{X}$ denote the group of entities
sharing the same preceding-level SID as $x$. We define
\begin{equation}
\mathcal{A}_x
=
\left\{
k_{x'}^{(N_q)}
\mid
x'\in\mathcal{G}_x\setminus\{x\}
\right\}
\end{equation}
as the set of final-level code indices already occupied within this
group. The replacement source is selected as:
\begin{equation}
x^{*}
=
\arg\max_{\substack{
x'\in\mathcal{X}\setminus\{x\}\\
k_{x'}^{(N_q)}\notin\mathcal{A}_x
}}
\cos\!\left(
\mathbf{e}_x^{\mathrm{coll}},
\mathbf{e}_{x'}^{\mathrm{coll}}
\right).
\end{equation}
We then replace $k_x^{(N_q)}$ with $k_{x^{*}}^{(N_q)}$ while keeping
all preceding code indices unchanged. This procedure is repeatedly
applied to the remaining conflicting entities, with the occupied set
updated after each replacement, until no SID collision remains.

\subsection{Progressive Multi-level LLM Alignment}
\label{sec:llm_alignment}

After obtaining SIDs, we incorporate them into the LLM vocabulary as randomly initialized new tokens and align them with the language space through supervised fine-tuning. Grounded in the collaborative semantic association of ``POI category $\rightarrow$ travel intent $\rightarrow$ behavioral pattern'', we construct three alignment datasets, $\mathcal{D}_{\text{index}}$, $\mathcal{D}_{\text{traj}}$, and $\mathcal{D}_{\text{rec}}$, for SID-level grounding, trajectory-level category-intent understanding, and generative next-POI prediction, respectively. Existing LLM-based generative recommendation approaches commonly adopt the multi-task fine-tuning (MFT) alignment paradigm proposed by LCRec~\cite{zheng2024LCRec}, where these datasets are mixed into $\mathcal{D}_{\text{align}} = \mathcal{D}_{\text{index}} \cup \mathcal{D}_{\text{traj}} \cup \mathcal{D}_{\text{rec}}$ and jointly optimized. However, such a strategy may suffer from sample imbalance caused by the larger recommendation-oriented dataset, and may ignore the inherent dependency among tasks. To address these issues, following the human learning process from basic concepts to complex problem solving, we propose a \textbf{Progressive Multi-level Training (PMT) alignment paradigm}. PMT organizes the three datasets into a three-stage curriculum: Stage 1 performs concept-level alignment using $\mathcal{D}_{\text{con}} = \mathcal{D}_{\text{index}}$, Stage 2 performs knowledge-level alignment using $\mathcal{D}_{\text{know}} = \mathcal{D}_{\text{index}} \cup \mathcal{D}_{\text{traj}}$, and Stage 3 performs task-level alignment using $\mathcal{D}_{\text{task}} = \mathcal{D}_{\text{rec}}$, ensuring that concept modeling precedes trajectory-level learning, while trajectory knowledge further supports the final recommendation task. 

\noindent\textbf{Stage 1: Concept-level Alignment.}
From a language perspective, each POI or user can be represented as a ``word'' formed by its corresponding SID, where a complete SID such as \texttt{<a\_1><b\_2><c\_3>} is composed of multiple discrete newly introduced tokens.
To enable the model to further compose these words into trajectory-level ``sentences'', the first step is to understand the semantic concept behind each word. Therefore, Stage 1 focuses on \textit{A. Explicit Dual Index-Language Alignment}, which maps the discrete SID tokens composing these words into meaningful semantic units in the LLM space. In this stage, we freeze the LLM backbone and optimize only the newly added SID embeddings using $\mathcal{D}_{\text{con}} = \mathcal{D}_{\text{index}}$, so that each SID obtains category-collaborative semantics before being composed into trajectory-level inputs. 
Freezing the backbone anchors the pretrained language space, forcing the randomly initialized SID embeddings to absorb category-collaborative semantics rather than allowing the LLM to adapt around semantically ungrounded tokens.

Specifically, $\mathcal{D}_{\text{index}}$ contains four concept-level alignment tasks. \textit{A.1 POI Category-Collaborative Reconstruction} generates the POI SID from its category description and multiple User SIDs that frequently visit the POI. Since a category alone cannot uniquely identify a POI, the associated User SIDs provide collaborative clues for distinguishing POIs within similar category regions. In practice, we use six associated User SIDs for each POI. \textit{A.2 POI Index to Category} recovers the corresponding category from a POI SID, encouraging the SID embedding to preserve interpretable category semantics. \textit{A.3 User Time-Category Preference to Index} generates the User SID from the user's long-term temporal-category preference profile $\mathcal{P}_u^{\mathrm{cat}}$. \textit{A.4 User Index to Time-Category Preference} recovers the user's preference description from a User SID. Together, these four tasks establish explicit bidirectional alignment between SIDs and language semantics.
This bidirectional supervision prevents one-way index memorization: reconstruction tasks teach the model to generate SIDs from semantic evidence, while inverse tasks require the same SIDs to recover their category or preference meanings.

Formally, for each input-output pair $(X,Y) \in \mathcal{D}_{\text{con}}$, where $X$ denotes the instruction and task-specific prompt, and $Y=(y_1,\ldots,y_{|Y|})$ denotes the target response, Stage 1 optimizes the autoregressive language modeling objective:
\begin{equation}
\mathcal{L}_{\text{con}}(\theta_1) =
-\sum_{(X,Y) \in \mathcal{D}_{\text{con}}}
\sum_{\ell=1}^{|Y|}
\log p_{\theta_0,\theta_1}(y_\ell \mid y_{<\ell}, X),
\end{equation}
where $\theta_1$ denotes the learnable embeddings of the newly added SID tokens, and $\theta_0$ denotes the frozen parameters of the LLM.

\noindent\textbf{Stage 2: Knowledge-level Alignment.}
After Stage 1, individual SIDs have been grounded as meaningful concept units in the LLM embedding space. However, since the LLM backbone is frozen in Stage 1, the model still needs to learn how to interpret the word-level knowledge conveyed by these SID-based concepts. Moreover, recommendation requires understanding trajectories formed by multiple check-ins, where POI SIDs and timestamps are composed into trajectory-level sentences. Therefore, Stage 2 aims to enable the full LLM to acquire both word-level SID knowledge and sentence-level trajectory knowledge. Meanwhile, we note that the essence of recommendation lies in modeling user preferences~\cite{Kang2018Sasrec}. Motivated by this, we construct an alignment dataset $\mathcal{D}_{\text{traj}}$ with two trajectory-level tasks for understanding users' historical behaviors.

To achieve knowledge-level alignment, Stage 2 jointly uses \textit{A. Explicit Dual Index-Language Alignment} and \textit{B. Trajectory Preference Understanding}. The former enables the LLM backbone to interpret the category or preference semantics conveyed by individual SIDs, while the latter teaches the model to understand user mobility behaviors from SID sequences. Specifically, $\mathcal{D}_{\text{traj}}$ contains two knowledge-level tasks. 
\textit{B.1 Trajectory Sequence Modeling} draws on the continued pre-training paradigm of LLMs~\cite{Liu2023survey} and performs self-supervised learning over historical trajectories composed of timestamp-SID pairs $(t_i, s_i)$, where $s_i$ denotes the SID of the POI visited at time $t_i$. This task encourages the model to capture temporal-category regularities, collaborative associations between time and SIDs, and sequential dependencies within user mobility trajectories. 
\textit{B.2 Trajectory Category--Intent Preference Description} asks the LLM to generate the user's temporal-category preference $\mathcal{P}_u^{\mathrm{cat}}$ and short-term mobility intent from the historical trajectory. This task requires the SID sequence to effectively substitute for the category sequence, enabling the LLM to comprehend the joint language and collaborative semantics within trajectories and accurately extract category-aware user intent. To obtain ground-truth preferences and intentions, we employ an LLM (\eg Deepseek-V3.2) to summarize the user's preferences and short-term intentions from the same historical trajectories.

Different from Stage 1, Stage 2 fine-tunes the full LLM on $\mathcal{D}_{\text{know}} = \mathcal{D}_{\text{index}} \cup \mathcal{D}_{\text{traj}}$, so that both the SID embeddings and the LLM backbone can adapt to knowledge-level semantic understanding. Formally, for each input-output pair $(X,Y)$ sampled from $\mathcal{D}_{\text{know}}$, starting from the model initialized by Stage 1 (i.e., $\theta_1$ for SID embeddings and $\theta_0$ for the rest), Stage 2 optimizes:
\begin{equation}
\mathcal{L}_{\text{know}}(\theta_2) =
-\sum_{(X,Y) \in \mathcal{D}_{\text{know}}}
\sum_{\ell=1}^{|Y|}
\log p_{\theta_2}(y_\ell \mid y_{<\ell}, X),
\end{equation}
where $\theta_2$ denotes all learnable parameters of the LLM, initialized from $\{\theta_0, \theta_1\}$.

\noindent\textbf{Stage 3: Task-level Alignment.}
After the LLM has acquired the necessary recommendation-related knowledge, Stage 3 focuses on \textit{C. Trajectory Sequence POI Prediction} and further fine-tunes the full model on $\mathcal{D}_{\text{task}} = \mathcal{D}_{\text{rec}}$ toward the final generative recommendation objective. Specifically, given a user's historical trajectory $\tau$, the model predicts the next POI $v_{n+1}$ that the user will visit at the next timestamp $t_{n+1}$. Formally, for each input-output pair $(X,Y) \in \mathcal{D}_{\text{task}}$, starting from the model initialized by Stage 2 (i.e., $\theta_2$), Stage 3 optimizes:
\begin{equation}
\mathcal{L}_{\text{task}}(\theta_3) =
-\sum_{(X,Y) \in \mathcal{D}_{\text{task}}}
\sum_{\ell=1}^{|Y|}
\log p_{\theta_3}(y_\ell \mid y_{<\ell}, X),
\label{eq:task_loss}
\end{equation}
where $\theta_3$ denotes all learnable parameters of the LLM, initialized from $\theta_2$.


Overall, PMT sequentially optimizes three stage-specific objectives,
where $\Theta^{(s)}$ denotes the model checkpoint after Stage~$s$:
\begin{equation}
\Theta^{(0)}
\xrightarrow{\mathcal{L}_{\mathrm{con}}}
\Theta^{(1)}
\xrightarrow{\mathcal{L}_{\mathrm{know}}}
\Theta^{(2)}
\xrightarrow{\mathcal{L}_{\mathrm{task}}}
\Theta^{(3)} .
\label{eq:pmt_total}
\end{equation}

It is worth emphasizing that each stage is trained with the final generative POI recommendation task as the validation objective. This design implicitly bridges the gap between textual and collaborative semantics, ensuring that the language semantics injected via prompts at each stage ultimately serve the recommendation goal. Moreover, it clarifies the role of each stage in the final task, which helps mitigate overfitting and guarantees the stability and robustness of the progressive training process. 
Detailed templates and prompt examples are provided in Appendices~\ref{appendix:user_preference_intent_generation} and~\ref{appendix:alignment_task_formats}.

%% file: sections/Experimental_Setting.tex
\begin{table}[!t]
\centering
\caption{Dataset statistics after preprocessing.
}
\label{tab:dataset_stats}
\resizebox{\columnwidth}{!}{%
\begin{tabular}{l c c c c c}
\toprule
\textbf{Dataset} & \textbf{Users} & \textbf{POIs} & \textbf{Trajs} & \textbf{Categories} & \textbf{Check-ins} \\
\midrule
NYC & 1,048 & 4,981 & 14,130 & 318 & 103,941 \\
TKY & 2,282 & 7,833 & 65,499 & 291 & 405,000 \\
CA  & 3,957 & 9,690 & 45,123 & 296 & 238,369 \\
\bottomrule
\end{tabular}%
}
\end{table}

\begin{table*}[!t]
\centering
\caption{Top-K recommendation performance on three datasets. SFT: supervised fine-tuning; RL: reinforcement learning.
Train/Test num: check-in limits for training/testing inputs.
$+$ indicates additional check-ins from the current trajectory.
Best overall results are in bold, and the strongest baseline results are underlined.
\textit{improve.} denotes the relative improvement over the strongest baseline available for each metric.
Methods marked with $^*$ can only produce top-1 recommendations due to their architectural constraints and therefore report Acc@1 only, with the remaining metrics marked as ``--''.}
\label{tab:top-k performance}
\resizebox{\textwidth}{!}{%
\begin{tabular}{@{}l *{16}{c}@{}}
\toprule
\multirow{2}{*}{Model}
& \multirow{2}{*}{Base LLM}
& \multirow{2}{*}{\makecell{Training\\paradigm}}
& \multirow{2}{*}{\makecell{Train\\num}}
& \multirow{2}{*}{\makecell{Test\\num}}
& \multicolumn{4}{c}{NYC}
& \multicolumn{4}{c}{TKY}
& \multicolumn{4}{c}{CA} \\
\cmidrule(lr){6-9}
\cmidrule(lr){10-13}
\cmidrule(lr){14-17}
& & & & &
Acc@1 & Acc@5 & Acc@10 & MRR &
Acc@1 & Acc@5 & Acc@10 & MRR &
Acc@1 & Acc@5 & Acc@10 & MRR \\
\midrule

PRME
& --- & --- & --- & ---
& 0.1159 & 0.2236 & 0.3105 & 0.1712
& 0.1052 & 0.2278 & 0.2944 & 0.1786
& 0.0521 & 0.1034 & 0.1425 & 0.1002 \\

LSTM
& --- & --- & --- & ---
& 0.1305 & 0.2719 & 0.3283 & 0.1857
& 0.1335 & 0.2728 & 0.3277 & 0.1834
& 0.0665 & 0.1306 & 0.1784 & 0.1201 \\

PLSP
& --- & --- & --- & ---
& 0.1917 & 0.3678 & 0.4523 & 0.2806
& 0.1889 & 0.3523 & 0.4150 & 0.2542
& 0.1072 & 0.2278 & 0.2995 & 0.1847 \\

STAN
& --- & --- & --- & ---
& 0.2231 & 0.4582 & 0.5734 & 0.3253
& 0.1963 & 0.3798 & 0.4464 & 0.2852
& 0.1104 & 0.2348 & 0.3018 & 0.1869 \\

GETNext
& --- & --- & --- & ---
& 0.2406 & 0.4815 & 0.5811 & 0.3528
& 0.1829 & 0.4045 & 0.4961 & 0.2853
& 0.1526 & 0.3278 & 0.3946 & 0.2364 \\

STHGCN
& --- & --- & --- & ---
& 0.2734 & 0.5361 & 0.6244 & 0.3915
& 0.2950 & 0.5207 & 0.5980 & 0.3986
& 0.1730 & 0.3529 & 0.4191 & 0.2558 \\

TPG
& --- & --- & --- & ---
& 0.2555 & 0.5005 & 0.5932 & 0.3669
& 0.1420 & 0.3631 & 0.4492 & 0.2436
& 0.1749 & 0.3285 & 0.3860 & 0.2479 \\

ROTAN
& --- & --- & --- & ---
& 0.3106 & 0.5281 & 0.6131 & 0.4104
& 0.2458 & 0.4626 & 0.5392 & 0.3475
& 0.2199 & 0.3718 & 0.4334 & 0.2931 \\

\midrule

LLM4POI
& LLaMA3.1-8B & SFT & $300+$ & $300+$
& 0.3372 & 0.3982 & 0.5010 & 0.3807
& 0.3035 & 0.3797 & 0.4474 & 0.3492
& 0.2065 & 0.2586 & 0.3109 & 0.2371 \\

GNPR-SID
& LLaMA3.1-8B & SFT & 50 & 50
& 0.3618 & 0.4472 & 0.5521 & 0.4133
& 0.3062 & 0.3930 & 0.4668 & 0.3662
& 0.2403 & 0.2846 & 0.3384 & 0.2559 \\

Refine-POI
& LLaMA3.1-8B & RL & $50+$ & $50+$
& 0.3638
& 0.6011
& 0.6696
& 0.4651
& 0.3214
& 0.5216
& 0.5881
& 0.4072
& 0.2174
& 0.3755
& 0.4308
& 0.2833 \\

Think2Go
& LLaMA3.1-8B & SFT+RL & 50 & 50
& 0.3847
& \underline{0.6287}
& \underline{0.6835}
& \underline{0.4663}
& 0.3239
& \underline{0.5461}
& \underline{0.6051}
& \underline{0.4147}
& 0.2583
& \underline{0.4190}
& \underline{0.4682}
& \underline{0.3167} \\

\midrule

GA-LLM$^*$
& LLaMA2-7B & SFT & $600+$ & $600+$
& 0.3919 & -- & -- & --
& 0.3482 & -- & -- & --
& 0.2566 & -- & -- & -- \\

Gwhere$^*$
& Qwen2.5-7B & SFT+RL & 50 & 50
& 0.4027 & -- & -- & --
& 0.3310 & -- & -- & --
& 0.2721 & -- & -- & -- \\

ROS$^*$
& Qwen3-4B & SFT+RL & 50 & 300
& \underline{0.4181} & -- & -- & --
& \underline{0.3527} & -- & -- & --
& \underline{0.2874} & -- & -- & -- \\

\midrule

AlignGPR
& Qwen2.5-1.5B & SFT & 50 & 50
& 0.4270
& 0.6655
& 0.7118
& 0.5267
& 0.3513
& 0.5512
& 0.6137
& 0.4370
& 0.2894
& 0.4490
& 0.4891
& 0.3569 \\

\multicolumn{1}{@{}l}{\textit{improve.}}
& --- & --- & --- & ---
& 2.13\% \textuparrow
& 5.85\% \textuparrow
& 4.14\% \textuparrow
& 12.95\% \textuparrow
& 0.40\% \textdownarrow
& 0.93\% \textuparrow
& 1.42\% \textuparrow
& 5.38\% \textuparrow
& 0.70\% \textuparrow
& 7.16\% \textuparrow
& 4.46\% \textuparrow
& 12.69\% \textuparrow \\

AlignGPR
& Qwen2.5-1.5B & SFT & 50 & 300
& \textbf{0.4464}
& \textbf{0.6696}
& \textbf{0.7214}
& \textbf{0.5409}
& \textbf{0.3763}
& \textbf{0.5894}
& \textbf{0.6486}
& \textbf{0.4668}
& \textbf{0.2915}
& \textbf{0.4689}
& \textbf{0.5157}
& \textbf{0.3652} \\

\multicolumn{1}{@{}l}{\textit{improve.}}
& --- & --- & --- & --- 
& 6.77\% \textuparrow
& 6.51\% \textuparrow
& 5.54\% \textuparrow
& 16.00\% \textuparrow
& 6.69\% \textuparrow
& 7.93\% \textuparrow
& 7.19\% \textuparrow
& 12.56\% \textuparrow
& 1.43\% \textuparrow
& 11.91\% \textuparrow
& 10.15\% \textuparrow
& 15.31\% \textuparrow \\

\bottomrule
\end{tabular}%
}
\end{table*}

\section{Experiments}
\subsection{Experimental Setting}
\label{Experimental Setting}

\noindent\textbf{Datasets.}
We evaluate our method on three benchmarks: Foursquare-NYC~\cite{Yang2014Four-square}, Foursquare-TKY~\cite{Yang2014Four-square}, and Gowalla-CA~\cite{Cho2011CA}. To ensure fair comparison, we strictly follow the preparation pipeline adopted by LLM4POI~\cite{Li2024LLM4POI}, Think2Go~\cite{Zhuang2026Think2Go}, and ROS\footnote{\url{https://github.com/amap-cvlab/Reasoning-Over-Space}}~\cite{Lv2026Ros}: (i) filter users and POIs with fewer than 10 check-ins; (ii) chronologically sort all check-in records and split them into trajectories using a 24-hour time-gap threshold, discarding trajectories with only one check-in; and (iii) split data into 80\%/10\%/10\% train/validation/test sets in temporal order while ensuring that validation/test users and POIs are observed in training. For each trajectory, the last check-in is taken as the target, and the previous check-ins form the input sequence. Table~\ref{tab:dataset_stats} reports the dataset statistics after preprocessing.

\noindent\textbf{Baselines.}
We compare AlignGPR with SOTA baselines from four categories: (1) \textit{Traditional methods:} PRME~\cite{Feng2015Markov} and LSTM~\cite{Hochreiter1997RNN}; (2) \textit{Neural-based methods:} STAN~\cite{Luo2021Stan}, GETNext~\cite{Yang2022GETNext}, STHGCN~\cite{Yan2023STHGCN}, TPG~\cite{luo2023TPG}, and ROTAN~\cite{feng2024rotan}; (3) \textit{LLM-based methods:} LLM4POI~\cite{Li2024LLM4POI}, GNPR-SID~\cite{Wang2025GNPR-SID}, Gwhere~\cite{Zhai2025CoAST}, GA-LLM~\cite{Liu2025GA-LLM} and ROS~\cite{Lv2026Ros}; and (4) \textit{Reasoning-based methods:} Refine-POI~\cite{Li2025Refine-POI} and Think2Go~\cite{Zhuang2026Think2Go}. More details about the baselines are provided in Appendix~\ref{app:baselines}.

\noindent\textbf{Evaluation metrics.}
We evaluate models' performance using Accuracy@K (Acc@K) and Mean Reciprocal Rank (MRR). More details about evaluation metrics are provided in Appendix~\ref{app:metrics}.

\subsection{Main Results}

Table~\ref{tab:top-k performance} reports the overall recommendation results on the three benchmark datasets. With training and test input limits of 50 and 300 check-ins, respectively, AlignGPR consistently outperforms all baselines across all reported metrics and datasets. Compared with ROTAN, AlignGPR improves MRR by 31.80\%, 34.33\%, and 24.60\% on NYC, TKY, and CA, respectively, demonstrating the advantage of using LLMs to model complex mobility patterns.

For top-$K$ recommendation, AlignGPR substantially outperforms GNPR-SID, with MRR gains of 27.44\%, 19.33\%, and 39.47\% on NYC, TKY, and CA, respectively, when both methods use training and test input limits of 50 check-ins. Under the same input limits, it also surpasses Think2Go, the strongest reasoning baseline reporting complete top-$K$ results, by 12.95\%, 5.38\%, and 12.69\% in MRR. Notably, these gains are achieved with only a 1.5B backbone and standard SFT, without relying on explicit reasoning generation or more complex optimization through reinforcement learning (RL). We attribute this advantage to PMT, which enables the LLM to jointly exploit explicit textual semantics related to POI categories and implicit collaborative signals for recommendation. These results suggest that effective semantic alignment can support strong POI recommendation performance even with a relatively small LLM and supervised fine-tuning.

For top-$1$ recommendation, AlignGPR outperforms Gwhere by 6.03\%, 6.13\%, and 6.36\% on NYC, TKY, and CA, respectively, with both methods using the 50/50 input limits. With both methods using the 50/300 input limits, it further surpasses ROS by 6.77\%, 6.69\%, and 1.43\%. These results demonstrate that AlignGPR not only retrieves the target POI among the top-$K$ candidates but also places it at the first position more accurately. We attribute this strong top-$1$ performance to the joint modeling of POI and user SIDs within the LLM, which aligns the category and collaborative semantics of POIs with the semantics of user preferences and enables precise matching between user intent and highly similar candidate POIs.

When the test input limit is increased from 50 to 300 check-ins, AlignGPR achieves further improvements across all metrics and datasets. This indicates that, although the model is trained on relatively short trajectories, it can generalize effectively to longer behavioral histories. We attribute this robustness to the SID design that integrates category semantics with collaborative signals and to progressive alignment across multiple levels, which help the model capture repetitive visitation patterns, collaborative semantics associated with POI categories, and fundamental mobility regularities from short trajectories.

\begin{table}[!t]
\centering
\caption{Ablation study on dual category-collaborative tokenization.
RID and SID denote random and semantic identifiers, respectively.}
\label{tab:ablation_tokenization}
\resizebox{\columnwidth}{!}{%
\begin{tabular}{c c c c c c c c}
\toprule
\multicolumn{2}{c}{Inputs} & \multicolumn{2}{c}{NYC} & \multicolumn{2}{c}{TKY} & \multicolumn{2}{c}{CA} \\
\cmidrule(lr){1-2} \cmidrule(lr){3-4} \cmidrule(lr){5-6} \cmidrule(lr){7-8}
POI & User & Acc@1 & MRR & Acc@1 & MRR & Acc@1 & MRR \\
\midrule
RID & RID & 0.3510 & 0.4302 & 0.3304 & 0.4132 & 0.2367 & 0.3034 \\
RID & SID & 0.3344 & 0.4094 & 0.3359 & 0.4245 & 0.2451 & 0.3186 \\
SID & RID & 0.4077 & 0.5146 & 0.3469 & 0.4342 & 0.2814 & 0.3502 \\
SID & SID & \textbf{0.4270} & \textbf{0.5267} & \textbf{0.3513} & \textbf{0.4370} & \textbf{0.2894} & \textbf{0.3569} \\
\bottomrule
\end{tabular}%
}
\end{table}

\subsection{Ablation Study}
We conduct ablation studies to evaluate the key components of AlignGPR from three perspectives: dual category-collaborative SID tokenization, concept- and knowledge-level alignment, and the comparison between MFT and PMT. All ablations are conducted under the 50-check-in setting.

\begin{table}[!t]
\centering
\caption{Ablation study on concept-level alignment and knowledge-level alignment.}
\label{tab:ablation_semantic}
\resizebox{\columnwidth}{!}{%
\begin{tabular}{cc cc cc cc}
\toprule
\multicolumn{2}{c}{Alignment} & \multicolumn{2}{c}{NYC} & \multicolumn{2}{c}{TKY} & \multicolumn{2}{c}{CA} \\
\cmidrule(lr){1-2} \cmidrule(lr){3-4} \cmidrule(lr){5-6} \cmidrule(lr){7-8}
Stage1 & Stage2 & Acc@1 & MRR & Acc@1 & MRR & Acc@1 & MRR \\
\midrule
$\times$ & $\times$ & 0.3510 & 0.4468 & 0.3260 & 0.4052 & 0.1651 & 0.2126 \\
$\surd$ & $\times$ & 0.4132 & 0.5145 & 0.3436 & 0.4300 & 0.2747 & 0.3450 \\
$\times$ & $\surd$ & 0.4084 & 0.4946 & 0.3371 & 0.4208 & 0.2709 & 0.3308 \\
$\surd$ & $\surd$ & \textbf{0.4270} & \textbf{0.5267} & \textbf{0.3513} & \textbf{0.4370} & \textbf{0.2894} & \textbf{0.3569} \\
\bottomrule
\end{tabular}%
}
\end{table}

\begin{figure}[!t]
    \centering
    \includegraphics[width=\columnwidth]{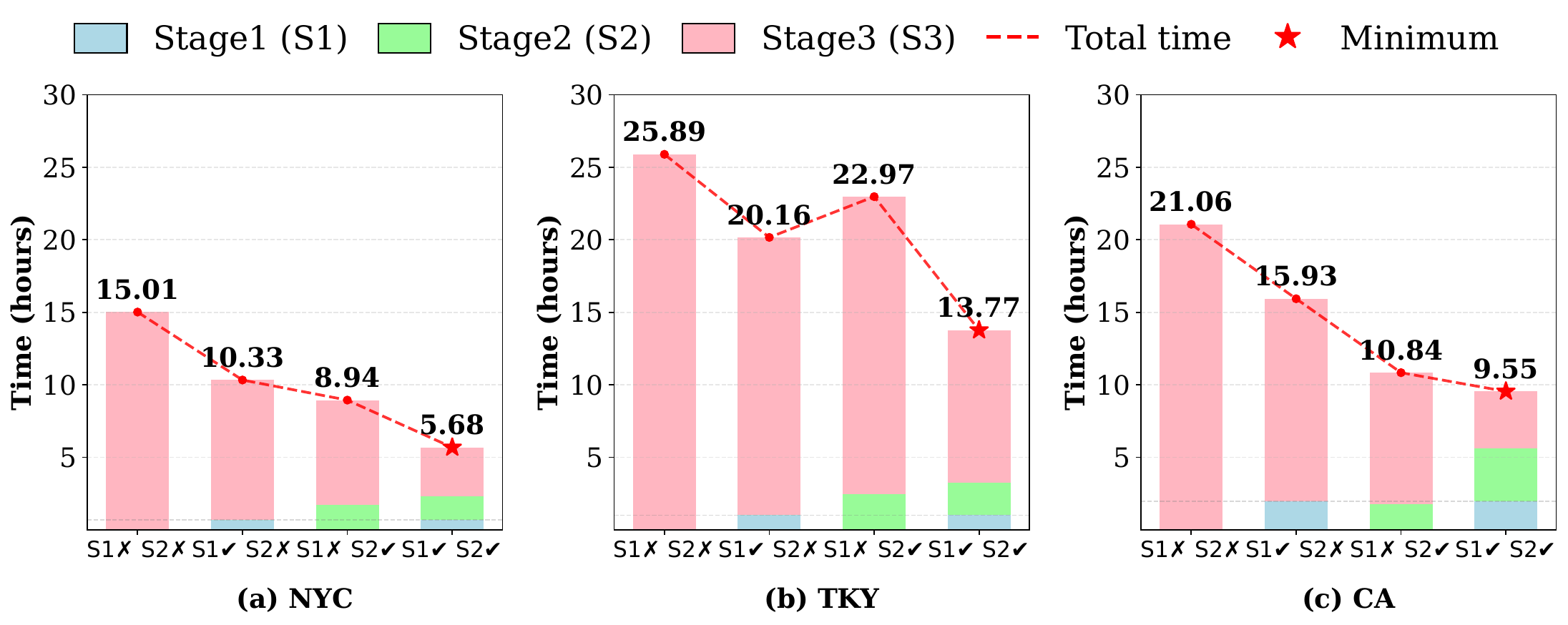}
    \caption{Training efficiency of AlignGPR with ablation of semantic alignment stages.}
    \label{fig:time_all_datasets}
\end{figure}

\noindent\textbf{Dual category-collaborative tokenization.}
Table~\ref{tab:ablation_tokenization} evaluates the contribution of dual category-collaborative tokenization. Replacing random POI identifiers (RIDs) with SIDs brings clear improvements, e.g., yielding a 16.2\% relative improvement in Acc@1 on NYC over the all-RID baseline, showing that category-aware POI SIDs provide useful semantic priors for inferring user intent.
In contrast, using user SIDs alone shows mixed effects, suggesting that the semantics of user preferences are difficult to exploit when POIs remain represented by RIDs.
The full dual-SID configuration achieves the best performance on all datasets, improving NYC Acc@1 by 21.6\% over the variant using RIDs for both POIs and users.
These results indicate that POI and user SIDs are complementary: POI SIDs provide explicit category-collaborative semantics, while user SIDs further integrate long-term preference patterns and provide compact, semantically rich representations that facilitate collaborative signal sharing across users.

\begin{table}[!t]
\centering
\caption{Effectiveness and training efficiency of alignment paradigms across datasets.}
\label{tab:alignment_paradigm}
\resizebox{\columnwidth}{!}{%
\begin{tabular}{l c c c c c c}
\toprule
\multirow{2}{*}{\textbf{\makecell{Alignment\\ Paradigm}}} & \multicolumn{2}{c}{\textbf{NYC}} & \multicolumn{2}{c}{\textbf{TKY}} & \multicolumn{2}{c}{\textbf{CA}} \\
\cmidrule(lr){2-3} \cmidrule(lr){4-5} \cmidrule(lr){6-7}
& \textbf{Acc@1} & \textbf{Time (h)} & \textbf{Acc@1} & \textbf{Time (h)} & \textbf{Acc@1} & \textbf{Time (h)} \\
\midrule
MFT              & 0.3939 & 20.07 & 0.3106 & 30.77 & 0.0870 & 26.01 \\
Stage1 + MFT     & 0.4250 & 13.12 & 0.3497 & 26.94 & 0.2849 & 20.72 \\
PMT              & \textbf{0.4270} & \textbf{5.68} & \textbf{0.3513} & \textbf{13.77} & \textbf{0.2894} & \textbf{9.55} \\
\bottomrule
\end{tabular}%
}
\end{table}

\noindent\textbf{Semantic alignment.}
Table~\ref{tab:ablation_semantic} and Figure~\ref{fig:time_all_datasets} demonstrate the necessity of arranging alignment tasks progressively rather than treating them as independent components. Concept-level alignment grounds newly introduced SIDs as meaningful semantic units, and knowledge-level alignment further connects these units to trajectory-level preference patterns. Notably, either alignment stage alone consistently improves performance over the no-alignment setting across all datasets, demonstrating the individual effectiveness of both stages. The best results are obtained when both stages are applied, showing that SID understanding should progress from concept-level to knowledge-level alignment. Although the full setting includes more alignment stages, it achieves the shortest total training time, indicating that PMT reduces the burden of the final recommendation stage by providing a structured semantic learning path before task-level optimization.

\begin{table}[!t]
\centering
\caption{User cold-start analysis on the NYC and TKY datasets.}
\label{tab:coldstart}
\resizebox{\columnwidth}{!}{%
\begin{tabular}{@{}l l c c c c c c@{}}
\toprule
\multirow{2}{*}{User Group}
& \multirow{2}{*}{Model}
& \multirow{2}{*}{\makecell{Train\\num}}
& \multirow{2}{*}{\makecell{Test\\num}}
& \multicolumn{2}{c}{NYC}
& \multicolumn{2}{c}{TKY} \\
\cmidrule(lr){5-6}
\cmidrule(lr){7-8}
& & & & Acc@1 & MRR & Acc@1 & MRR \\
\midrule

Inactive
& \multirow{3}{*}{STHGCN}
& \multirow{3}{*}{---}
& \multirow{3}{*}{---}
& 0.1460 & 0.2247 & 0.2164 & 0.3053 \\
Normal
& & &
& 0.3050 & 0.4265 & 0.2659 & 0.3596 \\
Very active
& & &
& 0.3085 & 0.4402 & 0.3464 & 0.4618 \\
\midrule

Inactive
& \multirow{3}{*}{LLM4POI}
& \multirow{3}{*}{$300+$}
& \multirow{3}{*}{$300+$}
& 0.2500 & 0.3043 & 0.2703 & 0.2930 \\
Normal
& & &
& 0.3689 & 0.3839 & 0.3014 & 0.3195 \\
Very active
& & &
& 0.3711 & 0.4079 & 0.3011 & 0.4092 \\
\midrule

Inactive
& \multirow{3}{*}{Refine-POI}
& \multirow{3}{*}{$50+$}
& \multirow{3}{*}{$50+$}
& 0.2832 & 0.3630 & 0.2655 & 0.3389 \\
Normal
& & &
& 0.3711 & 0.4602 & 0.2814 & 0.3630 \\
Very active
& & &
& 0.3860 & 0.5455 & 0.3768 & 0.4707 \\
\midrule

\multirow{2}{*}{Inactive}
& \multirow{6}{*}{AlignGPR}
& \multirow{6}{*}{50}
& 50
& \underline{0.3263} & \underline{0.4147}
& \underline{0.2929} & \underline{0.3732} \\
& & & 300
& \textbf{0.3444} & \textbf{0.4269}
& \textbf{0.3037} & \textbf{0.3866} \\

\multirow{2}{*}{Normal}
& & & 50
& \underline{0.4305} & \underline{0.5146}
& \underline{0.3157} & \underline{0.3952} \\
& & & 300
& \textbf{0.4677} & \textbf{0.5378}
& \textbf{0.3310} & \textbf{0.4178} \\

\multirow{2}{*}{Very active}
& & & 50
& \underline{0.4580} & \underline{0.5719}
& \underline{0.4066} & \underline{0.4986} \\
& & & 300
& \textbf{0.4762} & \textbf{0.5855}
& \textbf{0.4410} & \textbf{0.5391} \\

\bottomrule
\end{tabular}%
}
\end{table}

\noindent\textbf{MFT vs PMT.}
We further compare three alignment paradigms: MFT, Stage1+MFT, and our PMT. As shown in Table~\ref{tab:alignment_paradigm}, MFT suffers from inferior performance and higher training cost, especially on CA, suggesting that jointly optimizing SID understanding and next-POI prediction from scratch may introduce task interference under sparse trajectories. Adding Stage~1 before MFT clearly improves both accuracy and efficiency, confirming the importance of first grounding newly introduced SIDs as meaningful concept tokens. In contrast, PMT achieves the best Acc@1 and the shortest training time across all datasets,
demonstrating that ordered semantic alignment is more effective and efficient than jointly fine-tuning fully mixed semantic-alignment and recommendation tasks.

\subsection{User cold-start analysis}
Following prior next-POI recommendation studies~\cite{Yan2023STHGCN,Li2025Refine-POI}, we evaluate user cold-start generalization under trajectory sparsity, where users are grouped according to the number of their training trajectories. Specifically, we regard the lowest 30\% as Inactive, the highest 30\% as Very active, and the remaining users as Normal. We compare AlignGPR with STHGCN, LLM4POI, and Refine-POI, which incorporate collaborative signals to varying degrees and therefore provide relevant comparisons for user cold-start evaluation. As shown in Table~\ref{tab:coldstart}, AlignGPR outperforms all compared baselines across all user groups on both NYC and TKY under both the 50/50 and 50/300 settings. We attribute these gains to two complementary sources of collaborative semantics: user SIDs capture long-term preference patterns and facilitate collaborative signal transfer among behaviorally similar users, while category-collaborative semantics provide transferable category-level signals that remain informative under sparse observations. Their combination enables AlignGPR to infer user intent robustly from limited training trajectories. With a test input limit of 50 check-ins, relative MRR gains over the strongest baseline in each group are more pronounced for less active users, while extending the limit to 300 yields the largest additional relative gains for Normal users on NYC and Very active users on TKY.

\begin{table}[!t]
\centering
\caption{Zero-shot cross-domain performance (Acc@1). Each column denotes transfer from the source domain to the target domain without fine-tuning on the target domain.}
\label{tab:cross_domain}
\resizebox{\columnwidth}{!}{%
\begin{tabular}{l c c c c c c}
\toprule
Model & NYC-TKY & NYC-CA & TKY-NYC & TKY-CA & CA-NYC & CA-TKY \\
\midrule
LLM4POI   & 0.2594 & 0.1885 & 0.3463 & 0.1960 & 0.3344 & 0.2600 \\
GNPR-SID  & 0.2946 & 0.2297 & 0.3481 & 0.2342 & 0.3536 & 0.2869 \\
GA-LLM    & 0.3018 & 0.2053 & \underline{0.4059} & 0.2273 & 0.3670 & \underline{0.3065} \\
Think2Go  & \textbf{0.3183} & \textbf{0.2484} & 0.3523 & \underline{0.2545} & \underline{0.3677} & 0.3042 \\
AlignGPR  & \underline{0.3154} & \underline{0.2339} & \textbf{0.4118} & \textbf{0.2664} & \textbf{0.3973} & \textbf{0.3140} \\
\bottomrule
\end{tabular}}
\end{table}

\subsection{Cross-domain generalization analysis}
Table~\ref{tab:cross_domain} reports the zero-shot cross-domain transfer results, with AlignGPR using up to 300 historical check-ins at test time.
AlignGPR demonstrates strong cross-domain generalization across diverse urban environments, although its performance is relatively weaker when transferring from NYC, likely because of the dataset's smaller scale and more compact mobility patterns. These results suggest that AlignGPR captures transferable spatio-temporal regularities and latent collaborative associations rather than overfitting to patterns specific to the source domain, enabling the LLM to infer current mobility intent from POI visits and their timestamps despite differences in category semantics, POI spaces, and mobility distributions across cities. They also motivate future work on a unified tokenization scheme based on POI categories across domains to further improve transfer performance.

\subsection{Case Study}
\label{sec:case_study}
\label{appendix:case_study}

\begin{figure}[!t]
    \centering
    \includegraphics[width=1\linewidth]{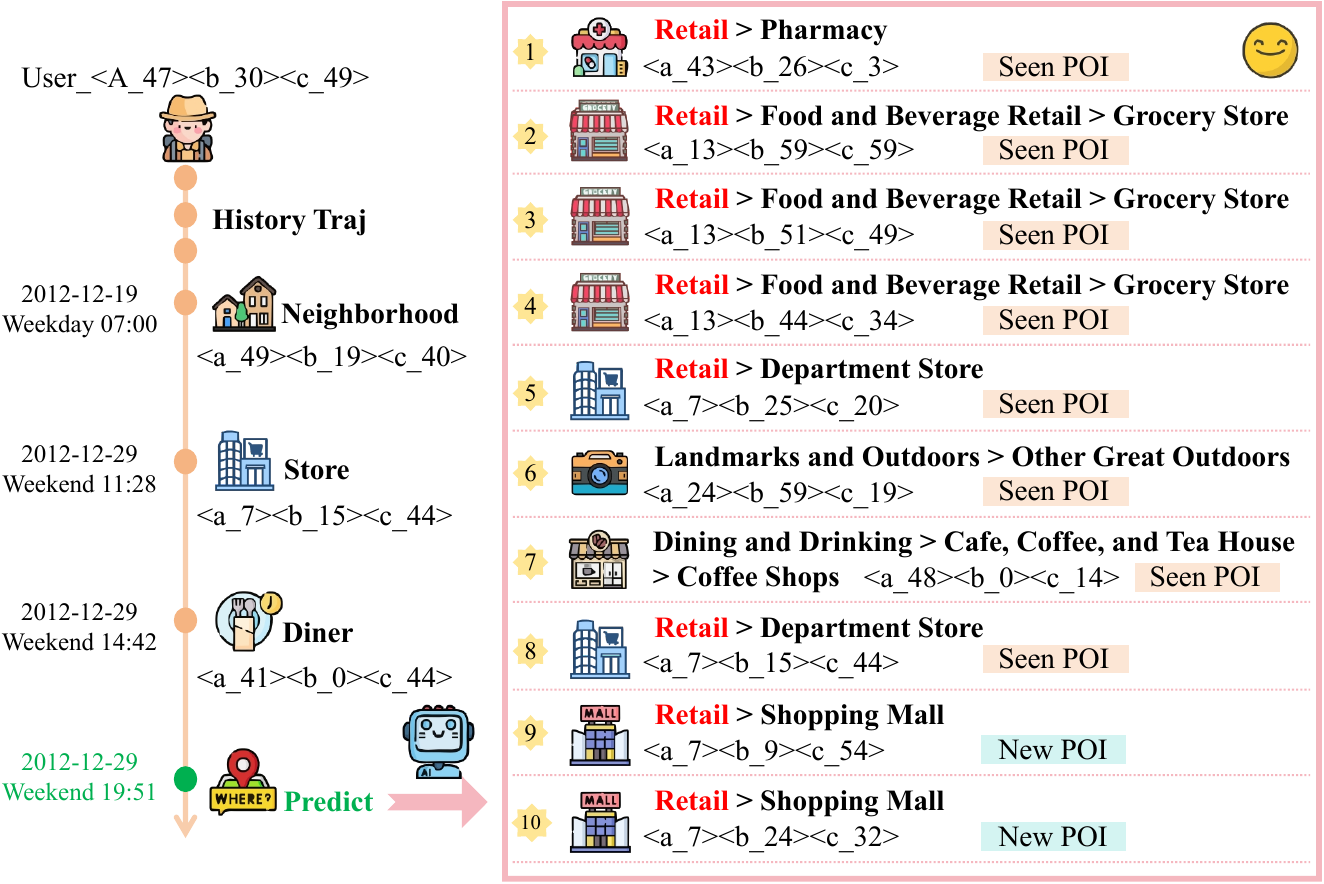}
    \caption{Case study on NYC. The left panel illustrates a recent segment of the user's trajectory, and the right panel shows the Top-10 generated POIs.
    }
    \label{fig:Case}
\end{figure}

Figure~\ref{fig:Case} provides an intuitive example of how AlignGPR makes preference-aligned recommendations from a user's recent trajectory. The Top-10 predictions show that AlignGPR balances revisitation and exploration. Eight predicted POIs have appeared in the user's observed history, while two have not. Notably, the two new POIs belong to \textit{Retail $>$ Shopping Mall}, which lies within the user's familiar category space. Although these two POIs do not appear in the user's observed history, their category does, illustrating how AlignGPR extends recommendations to new venues within familiar categories. The case also shows a pattern consistent with category-collaborative alignment. Although the recent trajectory includes non-retail activities, eight out of ten predictions fall into the Retail super-category. This suggests that AlignGPR considers the user's broader category preferences and mobility intent beyond individual activities in the recent trajectory. Thus, the generated recommendations connect the user's category preferences with specific POIs and remain coherent with the user's higher-level behavioral pattern.

%% file: sections/Conclusion.tex
\section{Conclusion}
\label{Conclusion}
We propose AlignGPR, a generative next POI recommendation framework that explicitly captures the POI Category → Mobility Intent → Behavioral Pattern progression, thereby deeply coupling category semantics with collaborative signals. By designing dual category-collaborative tokenization with collision resolution, joint semantic alignment tasks, and a progressive concept–knowledge–task training paradigm that naturally follows this progression, AlignGPR effectively bridges the semantic gap between textual knowledge and collaborative preferences. 
Extensive experiments on three real-world benchmarks demonstrate that AlignGPR achieves state-of-the-art accuracy and generalization while significantly reducing time and memory overhead.

%% file: sections/Appendix.tex
\section{Experimental Setting}
\label{app:experimental_setting}

\subsection{Evaluation Metrics}
\label{app:metrics}

For next-POI recommendation, we adopt Accuracy@K (Acc@K) and Mean Reciprocal Rank (MRR). Given a user $u$, let $\operatorname{rank}_u$ denote the rank position of the ground-truth next POI $p_u^{n+1}$ in the generated recommendation list. If the ground-truth POI is not retrieved, its contribution is set to 0. Acc@K is defined as
\begin{equation}
\operatorname{Acc@}K =
\frac{1}{|U|}
\sum_{u \in U}
\mathbf{1}\left(\operatorname{rank}_u \leq K\right),
\end{equation}
where $U$ is the set of evaluated users. MRR is defined as
\begin{equation}
\operatorname{MRR} =
\frac{1}{|U|}
\sum_{u \in U}
\frac{\mathbf{1}\left(\operatorname{rank}_u < \infty\right)}
{\operatorname{rank}_u}.
\end{equation}

For SID construction quality, we report Unique Category Prefix Rate (UCPR) and Collision Rate (CR). UCPR measures whether each two-level SID prefix corresponds to a unique fine-grained POI category. Let $\mathcal{X}$ be the set of distinct two-level SID prefixes. For each prefix $x \in \mathcal{X}$, its associated category set is
\begin{equation}
\mathcal{C}(x)
=
\left\{
\operatorname{cat}(i)
\,\middle|\,
(\mathbf{c}_i^{(1)}, \mathbf{c}_i^{(2)}) = x
\right\},
\end{equation}
where $\operatorname{cat}(i)$ denotes the fine-grained category of POI $i$. UCPR is defined as
\begin{equation}
\operatorname{UCPR}
=
\frac{1}{|\mathcal{X}|}
\sum_{x \in \mathcal{X}}
\mathbf{1}\left(|\mathcal{C}(x)| = 1\right).
\end{equation}

CR measures the proportion of POIs involved in full-SID collisions before collision handling. Let $P$ denote the POI set and $\mathcal{S}$ denote the set of distinct full SIDs. For each full SID $s \in \mathcal{S}$, let $\operatorname{freq}(s)$ be its frequency. CR is defined as
\begin{equation}
\operatorname{CR}
=
\frac{
\sum_{s \in \mathcal{S}}
\left(\operatorname{freq}(s)-1\right)
}{
|P|
}.
\end{equation}

\subsection{baseline}
\label{app:baselines}
\noindent\textbf{Neural-based methods.}
\textbf{STAN}~\cite{Luo2021Stan} is a spatio-temporal attention network that models spatial and temporal intervals between historical check-ins.
\textbf{GETNext}~\cite{Yang2022GETNext} enhances Transformer-based trajectory modeling with a global trajectory flow map, thereby incorporating collaborative transition signals among POIs.
\textbf{STHGCN}~\cite{Yan2023STHGCN} is a spatio-temporal hypergraph convolutional network that captures both inter-user and intra-user relationships for next-POI recommendation.
\textbf{TPG}~\cite{luo2023TPG} treats timestamps as prompts and integrates shifted geographic windows to enhance geography-aware location recommendation.
\textbf{ROTAN}~\cite{feng2024rotan} is a rotation-based temporal attention network that models time-specific user mobility patterns with rotary positional encoding.

\noindent\textbf{LLM-based methods.}
\textbf{LLM4POI}~\cite{Li2024LLM4POI} casts next-POI recommendation as a natural-language question answering task and adapts LLMs through supervised fine-tuning.
\textbf{GNPR-SID}~\cite{Wang2025GNPR-SID} introduces semantic IDs for generative next-POI recommendation, enabling the LLM to generate POIs as discrete semantic identifier sequences.
\textbf{Gwhere}~\cite{Zhai2025CoAST} is a cognitive-aligned spatio-temporal LLM framework that combines enriched pretraining, supervised fine-tuning, and reinforcement learning for next-POI prediction.
\textbf{GA-LLM}~\cite{Liu2025GA-LLM} is a geography-aware LLM method that injects GPS signals and transition patterns into the recommendation model.
\textbf{ROS}~\cite{Lv2026Ros} employs hierarchical spatial semantic identifiers and multi-task fine-tuning for generative next-POI recommendation, and further introduces spatial-guided reinforcement learning to optimize geographic consistency.

\noindent\textbf{Reasoning-based methods.}
\textbf{Refine-POI}~\cite{Li2025Refine-POI} applies reinforcement fine-tuning to large language models for top-$k$ next-POI recommendation.
\textbf{ROS$_{\mathrm{CoT}}$} additionally incorporates Mobility Chain-of-Thought reasoning to explicitly model intermediate mobility decisions before predicting the next POI.
\textbf{Think2Go}~\cite{Zhuang2026Think2Go} is a generative next-POI recommendation framework that combines supervised fine-tuning, reasoning enhanced through reinforcement learning, and test-time scaling to improve SID comprehension and recommendation performance.

\subsection{Implementation Details}
AlignGPR first converts POI categories and each user's long-term temporal-category preferences into dense embeddings. Given the difference in semantic richness, POI categories are embedded using the Qwen3-Embedding-0.6B encoder, while users' long-term temporal-category preferences are embedded using the Qwen3-Embedding-4B encoder, after which RQ-VAE performs residual quantization. For constructing the POI and user indices, the RQVAE is trained on a single GPU with a batch size of 2,048, a learning rate of $1 \times 10^{-3}$, and 10,000 epochs.
Considering data scale and cross-domain generalization, both RQ-VAEs use three residual quantization levels, each containing 64 codebook vectors of dimension 64, with codebook sizes following the public implementation of GNPR-SID\footnote{\url{https://github.com/wds1996/GNPR-SID}}~\cite{Wang2025GNPR-SID}.

For the SID–language alignment training, we adopt Qwen2.5-1.5B as the backbone model. The SID tokens are appended to the tokenizer vocabulary with randomly initialized embeddings. In stage1 (concept-level alignment), we freeze the backbone and only train the newly added SID tokens, using a learning rate of $5 \times 10^{-4}$, a total batch size of 1024, and training for 20 epochs on each dataset. In stage2 (knowledge-level alignment), we perform full-parameter supervised fine-tuning with a learning rate of $1 \times 10^{-4}$, a total batch size of 1024, and training for 10 epochs on each dataset. In stage3 (task-level alignment), we also perform full-parameter fine-tuning. Considering dataset scale, we set the learning rate to $4 \times 10^{-5}$ with a total batch size of 1024, and train for 10 epochs on NYC, 5 epochs on TKY, and 3 epochs on CA. For all stages, we employ the AdamW optimizer with a linear warmup of 20 steps. To mitigate overfitting and hyperparameter sensitivity, we apply an early stopping strategy that terminates training if the validation loss does not decrease for two consecutive evaluation steps, using a fixed evaluation frequency of every 1.25\% of an epoch across all datasets and stages. Each stage is initialized from the checkpoint with the lowest validation loss from the previous stage.
During inference, we evaluate over the entire POI set without candidate sampling, using constrained beam search with a beam size of 10 to generate recommendations. For each dataset, we select the checkpoint with the lowest validation loss for final evaluation.

For cross-domain generalization, we follow Think2Go~\cite{Zhuang2026Think2Go} and evaluate in a zero-shot setting. We first construct SIDs separately for each dataset, then directly evaluate the LLM trained on SID data from the source domain on SID test data from the target domain, without further LLM training.

For all baselines, we use the experimental settings and backbone models from their original papers and adopt officially released implementations whenever available. We report the better of our reproduced results and those reported in the original papers.

Following prior work \cite{Wang2025GNPR-SID,Li2025Refine-POI,Lv2026Ros} to maintain efficiency and enable fair comparison, we truncate the user’s historical trajectory to at most 50 check-ins during training. During evaluation, we report results under both the 50 check-ins setting and the full 300 check-ins setting. All experiments are conducted using 4 NVIDIA L40 48GB GPUs.

\section{More Ablation}
\label{appendix:ablation}

\subsection{SID Construction Ablation}
\label{app:sid_construction_ablation}

\begin{table}[!t]
\centering
\caption{Ablation of SID construction on NYC under the 50/50 setting.}
\label{tab:sid_construction_ablation}
\resizebox{\columnwidth}{!}{%
\begin{tabular}{lcc}
\toprule
Variant & Acc@1 & MRR \\
\midrule
Only collaborative embeddings & 0.3946 & 0.4625 \\
Unique suffix for collision handling & 0.4187 & 0.5223 \\
AlignGPR & \textbf{0.4270} & \textbf{0.5267} \\
\bottomrule
\end{tabular}%
}
\end{table}

Table~\ref{tab:sid_construction_ablation} examines the features used to construct SIDs and the strategy for resolving collisions on NYC under the 50/50 setting.

\noindent\textbf{Embedding features.}
Category features alone map POIs in the same category to the same SID before collision handling, making them indistinguishable through SIDs. Collaborative features alone encode collaborative similarity but lack explicit category semantics, hindering alignment between categories and mobility intents. As shown in Table~\ref{tab:sid_construction_ablation}, using only collaborative features reduces Acc@1 and MRR on NYC by 7.59\% and 12.19\%, respectively, compared with the full model. We attribute this decline to the difficulty of aligning SIDs constructed solely from collaborative features with the rich knowledge of category preferences in our alignment tasks, as their structure reflects collaborative similarity without explicitly encoding category semantics. These analyses and results support combining explicit category information with collaborative signals when constructing SIDs for next-POI recommendation.

\noindent\textbf{Collision handling.}
We compare our strategy with appending an additional unique suffix to each POI SID. Both strategies eliminate collisions and achieve the same UCPR of 0.7355. Our strategy improves Acc@1 and MRR by 1.98\% and 0.84\%, respectively, without increasing the length of POI SIDs. These results demonstrate the effectiveness of our collision handling strategy. By using collaborative similarity to modify only the final code index, it maintains unique POI identifiers while preserving the category structure encoded in the preceding SID levels.

\subsection{Knowledge-Level Alignment Ablation}
\label{appendix:knowledge_alignment_ablation}

\begin{table}[!t]
\centering
\caption{Performance of knowledge-level alignment ablation on NYC. A denotes word-level SID knowledge, and B denotes sentence-level trajectory knowledge.}
\label{tab:stage2_ablation_perf_nyc}
\resizebox{\columnwidth}{!}{%
\begin{tabular}{cccccc}
\toprule
A & B & Acc@1 & Acc@5 & Acc@10 & MRR \\
\midrule
$\times$ & $\times$ & $0.4132\,(\downarrow 3.33\%)$ & $0.6523\,(\downarrow 1.98\%)$ & $0.7076\,(\downarrow 0.59\%)$ & $0.5145\,(\downarrow 2.31\%)$ \\
$\surd$  & $\times$ & $0.4167\,(\downarrow 2.41\%)$ & $0.6537\,(\downarrow 1.80\%)$ & $0.7055\,(\downarrow 0.89\%)$ & $0.5163\,(\downarrow 1.97\%)$ \\
$\times$ & $\surd$  & $0.4146\,(\downarrow 2.90\%)$ & $0.6599\,(\downarrow 0.84\%)$ & $0.7104\,(\downarrow 0.19\%)$ & $0.5164\,(\downarrow 1.95\%)$ \\
$\surd$  & $\surd$  & $\mathbf{0.4270}$ & $\mathbf{0.6655}$ & $\mathbf{0.7118}$ & $\mathbf{0.5267}$ \\
\bottomrule
\end{tabular}%
}
\end{table}

\begin{table}[!t]
\centering
\caption{Training-time comparison of knowledge-level alignment ablation on NYC.}
\label{tab:stage2_ablation_time_nyc}
\begin{tabular}{ccccc}
\toprule
A & B & Stage2 Time & Stage3 Time & Total Time \\
\midrule
$\times$ & $\times$ & --      & 9:38:44 & 10:19:49 \\
$\surd$  & $\times$ & 09:36   & 6:35:07 & 7:25:48  \\
$\times$ & $\surd$  & 19:38   & 8:15:54 & 9:16:37  \\
$\surd$  & $\surd$  & 1:38:18 & 3:21:46 & \textbf{5:41:09} \\
\bottomrule
\end{tabular}
\end{table}

Tables~\ref{tab:stage2_ablation_perf_nyc} and~\ref{tab:stage2_ablation_time_nyc} analyze the role of knowledge-level alignment on NYC, where A denotes word-level SID knowledge and B denotes sentence-level trajectory knowledge. The results show that A and B are complementary: removing either component degrades performance, while using both achieves the best results across all metrics. Specifically, A grounds POI/User SIDs in the language space, whereas B strengthens trajectory-level preference modeling. Although the full Stage~2 setting requires additional alignment time, it reduces Stage~3 optimization time by 65.1\% and total training time by 45.0\% compared with omitting Stage~2. This suggests that knowledge-level alignment provides a better initialization for task-level fine-tuning, supporting the effectiveness of our progressive learning strategy and the organization of SID and trajectory knowledge.

\subsection{Optimization Dynamics under Semantic Alignment Ablations}
\label{app:stage3_optimization_dynamics}

\begin{figure}[!t]
    \centering
    \includegraphics[width=\columnwidth]{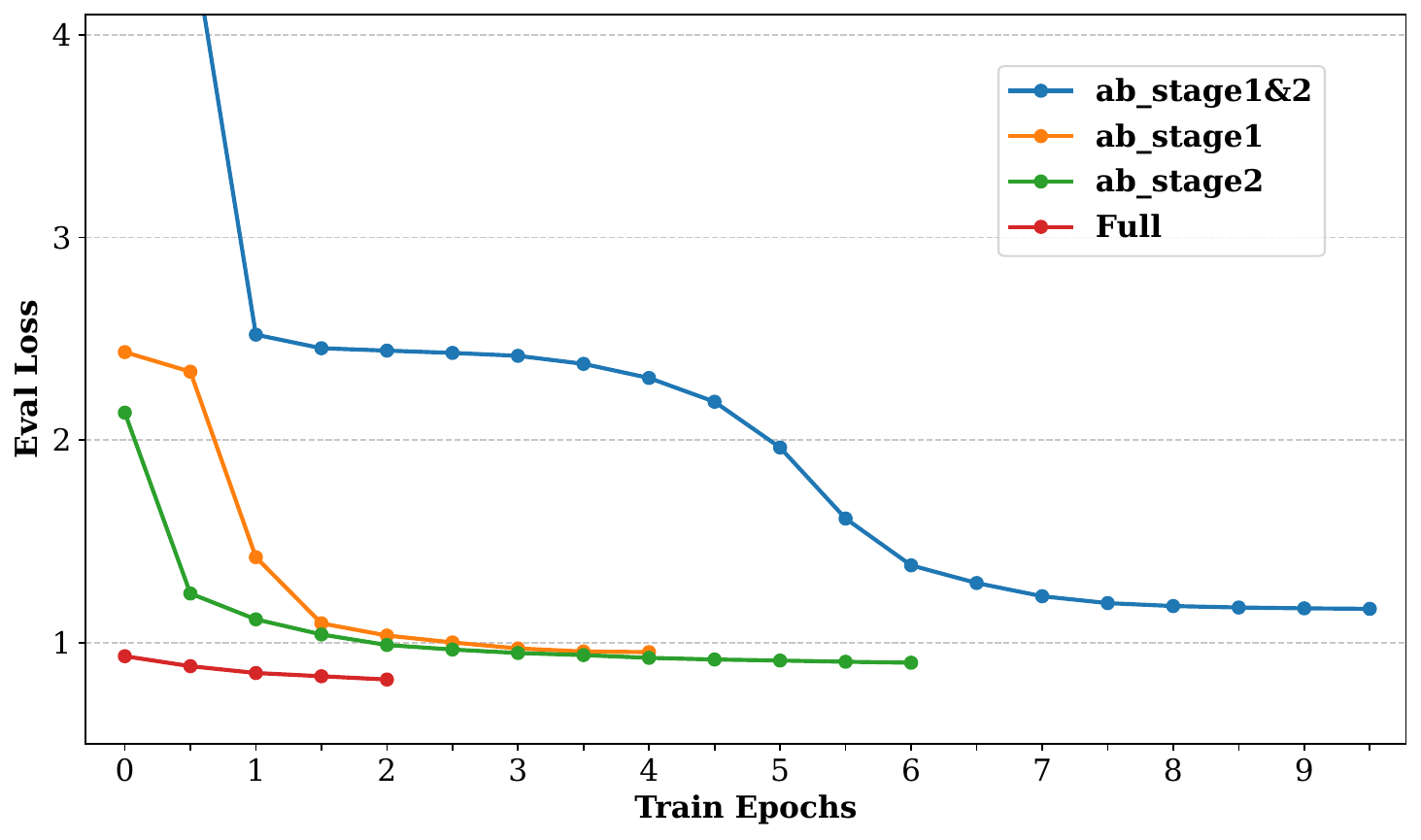}
    \caption{Evaluation loss of AlignGPR on NYC during Stage~3 under different semantic alignment ablations.}
    \label{fig:app_ablation_loss_curve}
\end{figure}

Figure~\ref{fig:app_ablation_loss_curve} shows the evaluation loss during Stage~3 on NYC under different semantic alignment ablations. The full model starts with the lowest evaluation loss and converges within the fewest epochs, indicating that concept-level and knowledge-level alignment provide effective semantic initialization before task-level recommendation learning. In contrast, removing either alignment stage leads to a higher initial loss and slower convergence, while removing both stages results in the largest optimization difficulty. These results further support that PMT improves not only recommendation accuracy but also the optimization efficiency of the recommendation.

\section{Model Efficiency Study}

\begin{table}[!t]
\centering
\caption{Efficiency comparison on NYC.}
\label{tab:efficiency}
\begin{tabular}{l c c c c c}
\toprule
\textbf{Model} & \makecell{\textbf{Test}\\\textbf{num}} & \textbf{GPU} & \makecell{\textbf{Total}\\\textbf{train}} & \makecell{\textbf{Test}\\\textbf{time}} & \makecell{\textbf{Peak}\\\textbf{memory}} \\
\midrule
Refine-POI  & $50+$ & 8 A100 & 13.5 h & 1.4 h  & 632 GB \\
Think2Go    & 50    & 5 L40S & 8.1 h  & 0.5 h  & 228 GB \\
AlignGPR    & 50    & 4 L40  & 5.7 h  & 0.06 h & 192 GB \\
\bottomrule
\end{tabular}
\end{table}

Computational efficiency is important for deploying generative POI recommendation models in practical scenarios. Table~\ref{tab:efficiency} compares the training time, inference time, and reported memory usage of AlignGPR, Refine-POI, and Think2Go on NYC under their respective hardware configurations and input settings. Refine-POI uses eight A100 GPUs, whereas Think2Go uses five L40S GPUs, including one for online sampling. With four L40 GPUs, a smaller 1.5B backbone, and standard SFT, AlignGPR has lower reported training time, inference time, and memory usage than the two 8B methods using RL. It generates 10 candidates with an average inference time of about 150 ms per test query. Together with the recommendation results, these findings indicate a favorable balance between recommendation accuracy and computational cost.

\section{Exploratory analysis.}
\begin{table}[!t]
\centering
\caption{Exploratory analysis on three datasets. Each model is trained on sequences of the most recent 50 check-ins. Prop.: proportion of test samples that are cold-start POIs.}
\label{tab:poi_coldstart}
\resizebox{\columnwidth}{!}{%
\begin{tabular}{@{}l c c c c c c c@{}}
\toprule
\multirow{2}{*}{Model} & \multirow{2}{*}{\makecell{Test\\num}} & \multicolumn{2}{c}{NYC} & \multicolumn{2}{c}{TKY} & \multicolumn{2}{c}{CA} \\
\cmidrule(lr){3-4} \cmidrule(lr){5-6} \cmidrule(lr){7-8}
& & Prop. & Acc@10 & Prop. & Acc@10 & Prop. & Acc@10 \\
\midrule
GNPR-SID & \multirow{2}{*}{50} & \multirow{2}{*}{23.70\%} & 0\% & \multirow{2}{*}{31.18\%} & 0\% & \multirow{2}{*}{43.99\%} & 0\% \\
AlignGPR & & & \textbf{2.04\%} & & \textbf{3.57\%} & & \textbf{4.68\%} \\
\midrule
ROS$_{\mathrm{CoT}}$      & \multirow{2}{*}{300} & \multirow{2}{*}{20.46\%} & 0\% & \multirow{2}{*}{22.94\%} & 0\% & \multirow{2}{*}{38.30\%} & 0\% \\
AlignGPR & & & \textbf{2.02\%} & & \textbf{4.28\%} & & \textbf{4.55\%} \\
\bottomrule
\end{tabular}}
\end{table}
This section evaluates whether a model can predict visits to POIs that have never appeared in the user's historical trajectory, which is particularly challenging because real-world mobility is dominated by revisitation behaviors. As shown in Table~\ref{tab:poi_coldstart}, AlignGPR is the first generative POI recommendation method to demonstrate that LLMs can reason about cold-start POIs through semantic tokens, highlighting the potential of LLMs to discover previously unvisited POIs that match users' preferences. We attribute the failure of GNPR-SID to its asynchronous optimization of SIDs without explicit semantics, while ROS$_{\mathrm{CoT}}$ pre-constrains the user intent space using historical mobility data and thus struggles to model generalizable preferences. In contrast, AlignGPR benefits from its category-collaborative SID design: category semantics support generalization to unseen POIs, while collaborative semantics enable fine-grained distinction through behavioral similarity. Notably, when the input trajectory length increases from 50 to 300, the proportion of cold-start POIs decreases because longer histories cover more visited POIs. Nevertheless, AlignGPR maintains comparable cold-start performance, indicating that it can still offer exploratory recommendations beyond the user's observed history even when more historical visits are provided.

\section{POI and User SID Analysis}

\begin{figure*}[!t]
\centering
\includegraphics[width=\textwidth]{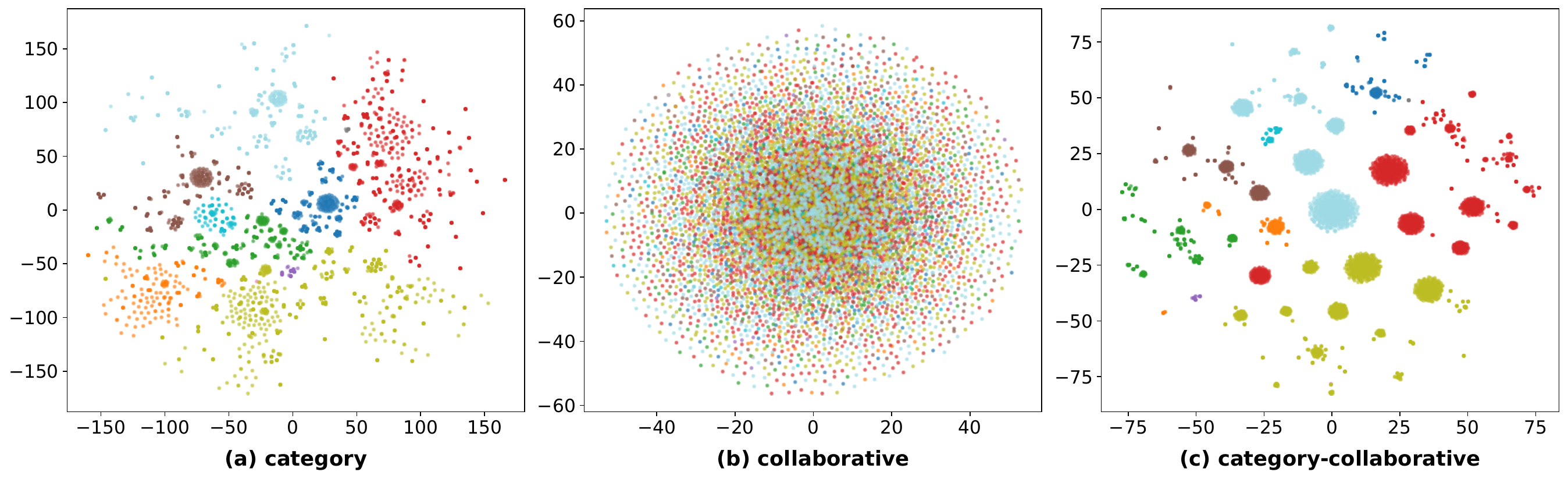}
\caption{Visualization of POI representations on TKY.}
\label{fig:poi_tsne}
\end{figure*}

\noindent\textbf{Visualization analysis of all POI representation}
Figure~\ref{fig:poi_tsne} compares the t-SNE projections of three POI representations on TKY. Category representations preserve basic semantic similarities, but POIs from the same fine-grained category remain loosely scattered, indicating that category semantics alone cannot distinguish their heterogeneous behavioral patterns. In contrast, collaborative representations exhibit a dense radial structure that captures interaction-derived regularities, yet POIs from different categories are extensively mixed due to the absence of explicit category semantics. Their integration produces more compact and clearly separated category-specific clusters while retaining fine-grained variations within each region. These results show that category semantics organize POIs according to mobility intent, whereas collaborative semantics discriminate behavioral patterns, providing an effective representation space for category-collaborative SID construction.

\begin{table}[!t]
\centering
\caption{Collision rates, UCPR, and codebook utilization of POI SIDs. L1--L3 denote the three residual quantization levels.}
\label{tab:sid_quality}
\resizebox{\columnwidth}{!}{%
\begin{tabular}{l c c c c c c}
\toprule
\multirow{2}{*}{\textbf{Dataset}}
& \multicolumn{2}{c}{\textbf{CR}}
& \multirow{2}{*}{\textbf{UCPR}}
& \multicolumn{3}{c}{\textbf{Codebook utilization}} \\
\cmidrule(lr){2-3} \cmidrule(lr){5-7}
& Before & After & & L1 & L2 & L3 \\
\midrule
NYC & 0.003 & 0.000 & 0.7355 & 100\% & 100\% & 100\% \\
TKY & 0.013 & 0.000 & 0.9300 & 100\% & 100\% & 100\% \\
CA  & 0.052 & 0.000 & 0.9233 & 100\% & 100\% & 100\% \\
\bottomrule
\end{tabular}%
}
\end{table}

\noindent\textbf{Quality Assessment of POI SIDs.}
Table~\ref{tab:sid_quality} shows that collision handling guided by collaborative semantics reduces the collision rate of full SIDs to zero on all three datasets while preserving substantial category consistency among POIs sharing a prefix. UCPR is particularly high on the larger TKY and CA datasets. Since collision handling modifies only the final code index, the preceding two-level prefixes remain unchanged and continue to encode the coarse semantic organization learned by residual quantization.

\begin{figure}[!t]
\centering
\includegraphics[width=\linewidth]{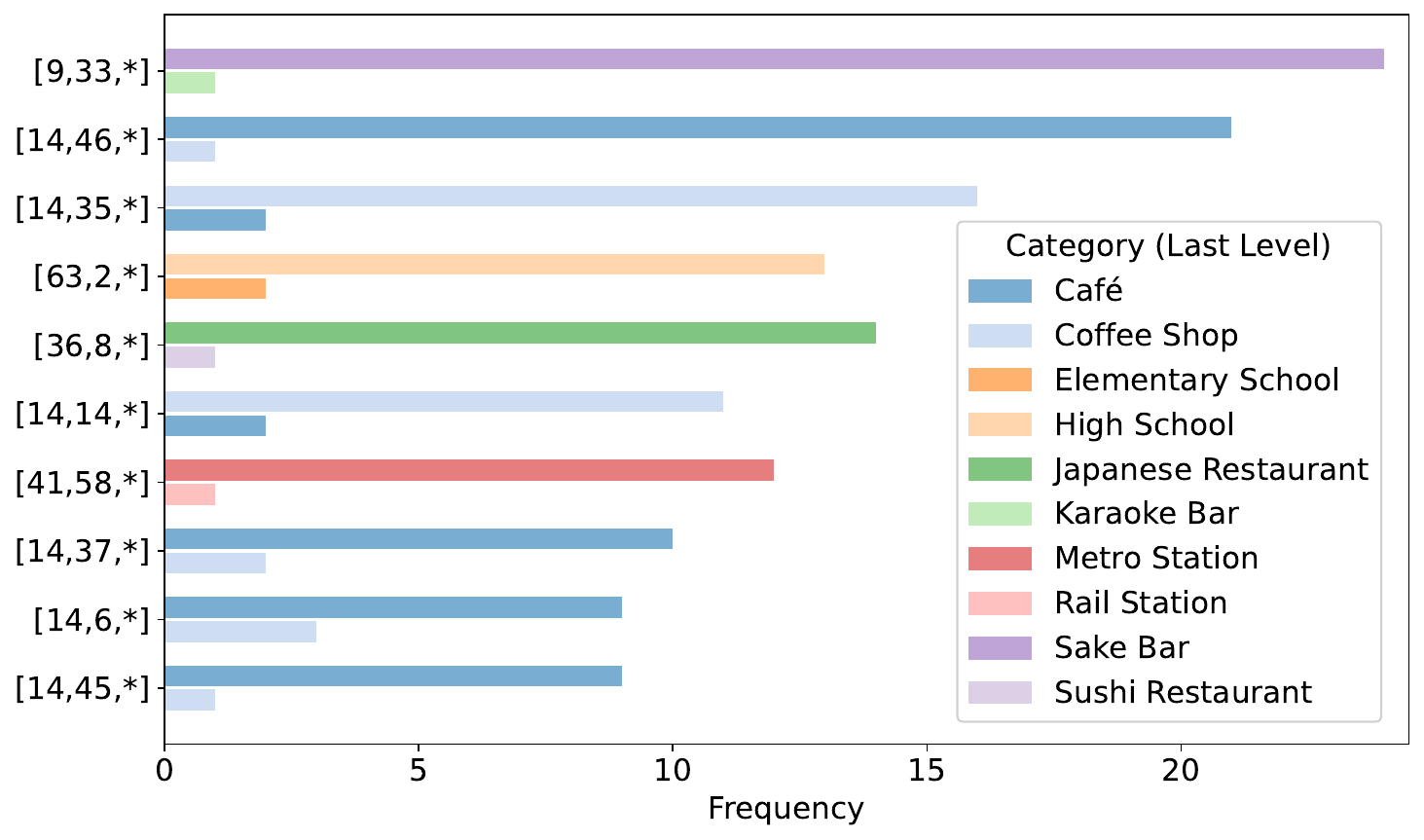}
\caption{Category composition of the 10 most frequent shared two-level POI SID prefixes on TKY.}
\label{fig:prefix_conflict_distribution}
\end{figure}

Figure~\ref{fig:prefix_conflict_distribution} further shows that each of the 10 most frequent shared two-level prefixes on TKY is dominated by a primary category, while the minor categories assigned to the same prefix are generally semantically or functionally related, such as cafés and coffee shops, metro and rail stations, or sake bars and karaoke bars. Such limited cross-category mixtures are reasonable rather than erroneous: closely related POI types often exhibit continuous semantic boundaries, overlapping functional roles, and similar mobility patterns in real-world scenarios. Therefore, sharing a coarse SID prefix reflects meaningful category-level and behavioral relatedness rather than random semantic mixing. Overall, our construction preserves interpretable semantic structure in the preceding SID levels while using the final level to ensure unique and collision-free full SIDs.

\begin{table}[!t]
\centering
\caption{Collision rates and codebook utilization of user SIDs. L1--L3 denote the three residual quantization levels.}
\label{tab:user_sid_quality}
\resizebox{\columnwidth}{!}{%
\begin{tabular}{l c c c c c}
\toprule
\multirow{2}{*}{\textbf{Dataset}}
& \multicolumn{2}{c}{\textbf{CR}}
& \multicolumn{3}{c}{\textbf{Codebook utilization}} \\
\cmidrule(lr){2-3} \cmidrule(lr){4-6}
& Before & After & L1 & L2 & L3 \\
\midrule
NYC & 0\%    & 0\% & 43\%  & 100\% & 100\% \\
TKY & 0\%    & 0\% & 100\% & 100\% & 100\% \\
CA  & 0.05\% & 0\% & 100\% & 100\% & 100\% \\
\bottomrule
\end{tabular}%
}
\end{table}

\noindent\textbf{Quality Assessment of User SIDs.}
Table~\ref{tab:user_sid_quality} reports the collision rates and codebook utilization of user SIDs, where utilization measures the proportion of codebook entries used at each quantization level. The collision rate is zero on NYC and TKY even before collision handling and decreases from 0.05\% to zero on CA. The second and third codebooks are fully utilized across all datasets, whereas first-level utilization is 43\% on NYC and 100\% on TKY and CA.

\noindent\textbf{Codebook Utilization and Cross-Domain Transfer.}
The POI RQ-VAE achieves full codebook utilization on all three datasets, ensuring that every constituent POI SID token has examples in the source domain for grounding its category and collaborative semantics. With a shared POI SID vocabulary across datasets, this provides support for zero-shot transfer at the vocabulary level. User SIDs encode personalized preferences, and we attribute NYC's lower utilization of the first codebook to its smaller user population and more limited coverage of preference patterns. When transferring from NYC to TKY or CA, some target user SID tokens therefore lack corresponding user examples during alignment in the source domain. This gap in semantic supervision further explains, in part, the weaker transfer performance from NYC in Table~\ref{tab:cross_domain}.

\section{Model Scale}
\label{app:model_scale}

\begin{table}[!t]
\centering
\caption{Model scale comparison on CA under the 50/50 setting.
Gwhere results are taken from~\cite{Zhai2025CoAST}.
Bold indicates the better result for each matched backbone.}
\label{tab:model_scale}
\begin{tabular}{l l c}
\toprule
\textbf{Model} & \textbf{Backbone} & \textbf{Acc@1} \\
\midrule
Gwhere & Qwen2.5-1.5B & 0.2585 \\
Gwhere & Qwen2.5-3B   & 0.2611 \\
Gwhere & Qwen2.5-7B   & 0.2721 \\
\midrule
AlignGPR & Qwen2.5-1.5B & \textbf{0.2894} \\
AlignGPR & Qwen2.5-3B   & \textbf{0.2901} \\
\bottomrule
\end{tabular}
\end{table}

Table~\ref{tab:model_scale} compares different Qwen2.5 backbone sizes on CA under the 50/50 setting. Increasing the backbone size of AlignGPR from 1.5B to 3B yields only a 0.24\% relative improvement in Acc@1. This modest gain is consistent with prior findings that scaling LLM recommenders can yield limited improvements in SID-based recommendation under SFT with fixed training data~\cite{Liu2026SIDScaling}. With the same 1.5B and 3B backbones, AlignGPR outperforms Gwhere by 11.95\% and 11.11\% in Acc@1, respectively. Even the 1.5B version of AlignGPR exceeds Gwhere-7B by 6.36\%. These results support our choice of the 1.5B backbone, which achieves strong recommendation performance with relatively few parameters.

We consider Gwhere's use of geographic knowledge and reinforcement learning for preference alignment~\cite{Zhai2025CoAST} an important factor in its performance gains from larger backbones. From a modeling perspective, AlignGPR provides a semantically aligned backbone that serves as a foundation for further optimization. In future work, we will investigate whether reinforcement learning and explicit reasoning can further exploit its learned mobility knowledge and the capacity of larger models.

\section{Alignment Task Formats}
\label{appendix:alignment_task_formats}

\subsection{Explicit Dual Index-Language Alignment}
\label{appendix:explicit_dual_index_language_alignment}

\noindent\textbf{1. POI Category-Collaborative Reconstruction.}
\label{appendix:poi_category_collaborative_reconstruction}

\begin{promptbox}
\sloppy
\noindent\textbf{Instruction:}\par
Answer the question about POI identification.

\noindent\textbf{User Input:}\par
Which POI has the hierarchical category:  Dining and Drinking $>$ Bar $>$ Sports Bar, and has been visited by these users: user\_\allowbreak<A\_31>\allowbreak<B\_40>\allowbreak<C\_42>, ...?

\noindent\textbf{Response:}\par
\texttt{<a\_3>\allowbreak<b\_63>\allowbreak<c\_57>}
\end{promptbox}

\noindent\textbf{2. POI Index to Category.}
\label{appendix:poi_index_to_category}

\begin{promptbox}
\sloppy
\noindent\textbf{Instruction:}\par
Answer the question about POI identification.

\noindent\textbf{User Input:}\par
What is the hierarchical category of POI \texttt{``<a\_41>\allowbreak<b\_0>\allowbreak<c\_44>''}?

\noindent\textbf{Response:}\par
Retail $>$ Pharmacy
\end{promptbox}

\noindent\textbf{3. User Time-Category Preference to Index.}
\label{appendix:user_time_category_preference_to_index}

\begin{promptbox}
\sloppy
\noindent\textbf{Instruction:}\par
Answer the question about user identification.

\noindent\textbf{User Input:}\par
Which user has the preference: The user exhibits a strong, stable preference for social and casual dining and drinking venues, ...?

\noindent\textbf{Response:}\par
\texttt{<A\_42>\allowbreak<B\_42>\allowbreak<C\_32>}
\end{promptbox}

\noindent\textbf{4. User Index to Time-Category Preference.}
\label{appendix:user_index_to_time_category_preference}

\begin{promptbox}
\sloppy
\noindent\textbf{Instruction:}\par
Answer the question about user identification.

\noindent\textbf{User Input:}\par
What is the preference of user\_\allowbreak<A\_42>\allowbreak<B\_42>\allowbreak<C\_32>?

\noindent\textbf{Response:}\par
The user exhibits a strong, stable preference for social and casual dining and drinking venues, ...
\end{promptbox}

\subsection{Trajectory Preference Understanding}
\label{appendix:trajectory_preference_understanding}

\noindent\textbf{1. Trajectory Sequence Modeling.}
\label{appendix:trajectory_sequence_modeling}

\begin{promptbox}
\sloppy
\noindent\textbf{Serialized Training Sequence:}\par
The POI visit history of user\_\allowbreak<A\_42>\allowbreak<B\_42>\allowbreak<C\_32> is as follows:\allowbreak
at 2012-04-13 Weekday 21:11,\allowbreak
visit <a\_60>\allowbreak<b\_15>\allowbreak<c\_12>;\allowbreak
... ;\allowbreak
at 2012-07-04 Weekday 15:04,\allowbreak
visit <a\_30>\allowbreak<b\_46>\allowbreak<c\_14>.
\end{promptbox}

\noindent\textbf{2. Trajectory Category-Intent Preference Description.}
\label{appendix:trajectory_category_intent_preference}

\begin{promptbox}
\sloppy
\noindent\textbf{Instruction:}\par
Analyze the user's POI visit history, provide insights about their preferences.

\noindent\textbf{User Input:}\par
Based on the user\_\allowbreak<A\_42>\allowbreak<B\_42>\allowbreak<C\_32>'s trajectory: visit <a\_60>\allowbreak<b\_15>\allowbreak<c\_12> at 2012-04-13 Weekday 21:11, ... Can you provide insights about the user\_\allowbreak<A\_42>\allowbreak<B\_42>\allowbreak<C\_32>'s preferences?

\noindent\textbf{Response:}\par
Based on the user's trajectory, the user shows a strong and stable preference for social food-and-drink scenarios, ...
\end{promptbox}

\subsection{Trajectory Sequence POI Prediction}
\label{appendix:trajectory_sequence_poi_prediction}

\begin{promptbox}
\sloppy
\noindent\textbf{Instruction:}\par
Can you predict the next possible POI that the user may visit at the specified time based on their POI visit history?

\noindent\textbf{User Input:}\par
Based on the user\_\allowbreak<A\_42>\allowbreak<B\_42>\allowbreak<C\_32>'s POI visit history:  at 2012-04-13 Weekday 21:11, visit <a\_60>\allowbreak<b\_15>\allowbreak<c\_12>; ... ;  at 2012-07-04 Weekday 15:04, visit <a\_30>\allowbreak<b\_46>\allowbreak<c\_14>. Can you predict the next POI that the user\_\allowbreak<A\_42>\allowbreak<B\_42>\allowbreak<C\_32> may visit at 2012-07-04 Weekday 21:00?

\noindent\textbf{Response:}\par
\texttt{<a\_31>\allowbreak<b\_24>\allowbreak<c\_62>}
\end{promptbox}

\section{User Preference and Trajectory Intent Generation Template}
\label{appendix:user_preference_intent_generation}

We use the following unified prompt template to generate long-term user preference profiles and trajectory-level intent descriptions.

\begin{promptbox}
\sloppy
\noindent\textbf{Prompt Template:}\par
Based on the user's hierarchical POI category visit history in
\texttt{\{Dataset\}}:

\texttt{\{poi\_seq\}}

(Each entry: ``visit [category] at [timestamp]'')

Conduct a comprehensive \textbf{preference (intent)} analysis covering the following core dimensions:

\begin{itemize}
    \item Long-term stable preferences and recurring patterns: frequently visited categories, consistent temporal habits, including weekday/weekend differentiation, intra-day time-slot preferences, and visit-frequency regularity, as well as core persistent demands over the entire period.

    \item Behavioral evolution and dynamic trends: shifts in demand orientation, phased changes in visit patterns, adjustments to temporal regularities, periodic fluctuations, and sequential patterns between categories, such as frequent transitions from one category to another, throughout the chronological sequence.

    \item Recent tendencies and preference deviations: emerging visit inclinations, newly formed habits, and significant deviations from long-term patterns in the most recent period.

    \item Core decision-making drivers: key underlying factors that dominate category selection, focusing on the intrinsic logic and core demands guiding the user's choices.
\end{itemize}

Provide a concise and general summary in a single paragraph, written in the third person and focusing on the core factors. The summary must use plain text only, without bullet points, numbered lists, headings, bold text, or other special formatting. Specific hierarchical POI categories may be mentioned when they are important for illustrating the user's preferences or trajectory intent. Start the summary with ``The user's preferences: The user'' (``Based on the user's trajectory, the user'' for trajectory-intent generation).
\end{promptbox}

\section{Sequential Data Augmentation}
\label{sec:data_augmentation}

Sequential data augmentation is widely adopted in non-reasoning sequential recommendation tasks, including SID-based generative recommendation\footnote{\url{https://github.com/RUCAIBox/LC-Rec}}%
\footnote{\url{https://github.com/AkaliKong/MiniOneRec}}%
\footnote{\url{https://github.com/ywc666/Align3GR}}
and Transformer-based POI recommendation%
\footnote{\url{https://github.com/songyangco/GETNext}}.

Given the chronologically ordered trajectory $\tau = \langle p_i \rangle_{i=1}^{n}$ of user $u$ and its corresponding timestamp sequence $T_u = \{t_1, t_2, \ldots, t_n\}$, we define the trajectory and timestamp prefixes preceding the $j$-th check-in as
\begin{equation}
\begin{aligned}
    \tau^{<j}
    &=
    \left(
        p_1, p_2, \ldots, p_{j-1}
    \right),
    \\
    T_u^{<j}
    &=
    \left(
        t_1, t_2, \ldots, t_{j-1}
    \right),
\end{aligned}
    \label{eq:trajectory_prefix}
\end{equation}
where $2 \leq j \leq n$. Let $v_j$ denote the POI associated with the check-in record $p_j$, and let
\begin{equation}
    Y_j
    =
    \left[
        y_{j}^{1},
        y_{j}^{2},
        \ldots,
        y_{j}^{L}
    \right]
    \in \mathcal{Y}^{L}
    \label{eq:target_poi_token_sequence}
\end{equation}
be the token sequence representing $v_j$, where $L$ is the fixed SID depth. For each non-initial position $j$, the model input is constructed as
\begin{equation}
\begin{aligned}
    X_j
    &=
    \operatorname{Concat}
    \left(
        I,\,
        \tau^{<j},\,
        T_u^{<j},\,
        t_j
    \right),
    \\
    &\qquad 2 \leq j \leq n,
\end{aligned}
    \label{eq:augmented_poi_input}
\end{equation}
where $I$ denotes the task instruction and $t_j$ is the target time of the next check-in relative to the prefix $\tau^{<j}$.

The all-prefix augmentation strategy treats every non-initial check-in as a prediction target, thereby constructing the following set of training instances:
\begin{equation}
    \mathcal{D}_u^{\mathrm{aug}}
    =
    \left\{
        \left(
            X_j, Y_j
        \right)
        \,\middle|\,
        2 \leq j \leq n
    \right\}.
    \label{eq:all_prefix_poi_augmentation}
\end{equation}
Accordingly, a trajectory containing $n$ check-ins provides $n-1$ supervision signals over user preference states at different stages. These prefix--target pairs may either be explicitly materialized as independent training instances or equivalently optimized using position-wise next-step supervision over a shifted trajectory.

In contrast, without sequential augmentation, each trajectory typically contributes only the final prefix--target instance:
\begin{equation}
    \mathcal{D}_u^{\mathrm{last}}
    =
    \left\{
        \left(
            X_n, Y_n
        \right)
    \right\}.
    \label{eq:last_target_poi_training}
\end{equation}
Consequently, the empirical training distribution becomes disproportionately concentrated on long trajectories and terminal transitions, biasing the model toward memorizing late-stage mobility patterns rather than capturing evolving user-specific preferences, and thereby weakening personalized POI recommendation for shorter histories and unseen mobility states.

For our Stage~3, we follow the same prefix-target construction while introducing a length-aware cap on the number of augmented instances. Trajectories in POI recommendation are substantially longer than the item sequences commonly used in conventional recommendation benchmarks. Applying exhaustive all-prefix augmentation would therefore produce $O(n)$ instances and $O(n^2)$ cumulative context tokens for a trajectory of length $n$, resulting in excessive training overhead.

To balance supervision coverage and computational cost, we retain at most the last $K$ valid prediction positions of each complete trajectory. Let
\begin{equation}
    K_u = \min\left(K, n - 1\right)
    \label{eq:adaptive_augmentation_size}
\end{equation}
denote the number of augmented instances generated for user $u$. The corresponding target-position set is defined as
\begin{equation}
    \mathcal{J}_u^{K}
    =
    \left\{
        j
        \,\middle|\,
        n - K_u + 1 \leq j \leq n
    \right\}.
    \label{eq:capped_target_positions}
\end{equation}
We then construct the Stage~3 training instances as
\begin{equation}
    \mathcal{D}_{u}^{\mathrm{cap}}
    =
    \left\{
        \left(X_j, Y_j\right)
        \,\middle|\,
        j \in \mathcal{J}_u^{K}
    \right\},
    \label{eq:capped_all_prefix_augmentation}
\end{equation}
where duplicate instances with identical user, trajectory prefix, timestamp prefix, target POI, and target time are removed.

This strategy applies exhaustive all-prefix augmentation to short trajectories satisfying $n-1 \leq K$, while restricting long trajectories to their latest $K$ prefix--target pairs. It therefore preserves dense supervision for short and sparse trajectories, while preventing long trajectories from disproportionately increasing the training corpus and dominating Stage~3 optimization.
Specifically, considering the maximum number of POIs allowed in each training trajectory and the statistics on trajectory lengths of the three datasets, we set $K=20$ for all experiments.